\documentclass{ws-ijmpa}

\usepackage{latexsym}
\usepackage{amsfonts}
\usepackage{amsmath}
\usepackage{amssymb}
\usepackage{revsymb}
\usepackage{bm}
\usepackage{color}

\usepackage{dcolumn} 
\usepackage[super,compress]{cite}

\newcolumntype{d}{D{.}{.}{0}}
\newlength{\textlength}
\newlength{\overlinelength}

\numberwithin{equation}{section}

\begin{document}

\newcommand{\be}{\begin{equation}}
\newcommand{\ee}{\end{equation}}
\newcommand{\bea}{\begin{eqnarray}}
\newcommand{\eea}{\end{eqnarray}}
\newcommand{\ack}[1]{[{\bf Pfft!: #1}]}

\newcommand{\beq}{\begin{equation}}
\newcommand{\eeq}{\end{equation}}
\newcommand{\beqn}{\begin{eqnarray}}

\newcommand{\eeqn}{\end{eqnarray}}
\newcommand{\pa}{\partial}
\newcommand{\osigma}{\overline{\sigma}}
\newcommand{\orho}{\overline{\rho}}
\newcommand{\de}{\mathtt{z}} 
\newcommand{\ftt}{{f_{tt}(T)}}
\newcommand{\Ee}{\mathrm{e}}
\newcommand{\ii}{\mathrm{i}}
\newcommand{\dif}{\mathrm{d}}
\newcommand{\bra}[1]{\langle #1 |}
\newcommand{\ket}[1]{| #1 \rangle}
\newcommand{\bbra}[1]{\bigl\langle #1 \bigr|}
\newcommand{\bket}[1]{\bigl| #1 \bigr\rangle}
\newcommand{\Bbra}[1]{\Bigl\langle #1 \Bigr|}
\newcommand{\Bket}[1]{\Bigl| #1 \Bigr\rangle}
\newcommand{\vev}[1]{\langle #1 \rangle}

\renewcommand{\slash}[2]{#1\hskip #2 em/}
\newcommand{\kslash}{\slash{k}{-0.5}}
\newcommand{\pslash}{\slash{p}{-0.5}}
\newcommand{\qslash}{\slash{q}{-0.5}}
\newcommand{\Pslash}{\slash{P}{-0.5}}
\newcommand{\dslash}{\slash{\partial}{-0.5}}

\newcommand{\eref}[1]{eq.\ (\ref{eq:#1})}
\def\NPB{{\it Nucl. Phys. }{\bf B}}
\def\PL{{\it Phys. Lett. }}
\def\PRL{{\it Phys. Rev. Lett. }}
\def\PRD{{\it Phys. Rev. }{\bf D}}
\def\CQG{{\it Class. Quantum Grav. }}
\def\JMP{{\it J. Math. Phys. }}
\def\SJNP{{\it Sov. J. Nucl. Phys. }}
\def\SPJ{{\it Sov. Phys. J. }}
\def\JETPL{{\it JETP Lett. }}
\def\TMP{{\it Theor. Math. Phys. }}
\def\IJMPA{{\it Int. J. Mod. Phys. }{\bf A}}
\def\MPL{{\it Mod. Phys. Lett. }}
\def\CMP{{\it Commun. Math. Phys. }}
\def\AP{{\it Ann. Phys. }}
\def\PR{{\it Phys. Rep. }}

\newcommand{\myfig}[3]{
	\begin{figure}[ht]
	\centering
	\includegraphics[width=#2cm]{#1}\caption{#3}\label{fig:#1}
	\end{figure}
	}
\newcommand{\littlefig}[2]{
	\includegraphics[width=#2cm]{#1}}
\newcommand{\1}{{\rm 1\hspace*{-0.4ex}%
\rule{0.1ex}{1.52ex}\hspace*{0.2ex}}}

\newcommand{\nn}{\nonumber}

\hyphenation{Min-kow-ski}
\hyphenation{cosmo-logical}
\hyphenation{holo-graphy}
\hyphenation{super-symmetry}
\hyphenation{super-symmetric}

\centerline{\bf ON NON-EQUILIBRIUM PHYSICS AND STRING THEORY}\vskip0.25cm
\centerline{{\footnotesize NICKOLAS GRAY\footnote{nickgray@vt.edu}, DJORDJE MINIC\footnote{dminic@vt.edu} and 
MICHEL PLEIMLING\footnote{Michel.Pleimling@vt.edu}
}}
\centerline{\footnotesize \it Department of Physics, Virginia Polytechnic Institute and State University,}
\centerline{\footnotesize \it Blacksburg, VA 24061, U.S.A.}


\begin{abstract}
In this article we review the relation between string theory and non-equilibrium physics based on our previously published work. First we explain why a theory of quantum gravity and non-equilibrium statistical physics should be related in the first place. Then we present the necessary background from the recent research in non-equilibrium physics. The review discusses the relationship of string theory and aging phenomena, as well as the connection between AdS/CFT correspondence and the Jarzynski identity. We also discuss the emergent symmetries in fully developed turbulence and the corresponding non-equilibrium stationary states. Finally we outline a larger picture regarding the relationship between non-perturbative quantum gravity and non-equilibrium statistical physics. This relationship can be understood as a natural generalization of the well-known Wilsonian relation between local quantum field theory and equilibrium statistical physics of critical phenomena. According to this picture the AdS/CFT duality 
is just an example of a more general connection between non-perturbative quantum gravity and non-equilibrium physics. In the appendix of this review we discuss a new kind of complementarity 
between thermodynamics 
and statistical physics which should be important in the context of black hole complementarity.
\end{abstract}

\pagestyle{plain}


\section{Outline of the Review}
Quantum gravity and non-equilibrium physics present two outstanding problems of contemporary physics.
In this article we review the surprising relation between string theory, viewed as a consistent theory of quantum gravity and matter, and non-equilibrium physics. We follow our publications, Refs. \refcite{age1,age2,age3,jarzads,dmturb}. (See also Refs. \refcite{otherpeople,otherpeople2}.) In section 2 we explain why a theory of quantum gravity and non-equilibrium statistical physics should be related in the first place. 
Then in section 3 we present the necessary background from the recent research in non-equilibrium physics. In section 4 we review the relationship of string theory and aging phenomena
as presented in Refs. \refcite{age1,age2,age3}. In section 5 we review the connection between the AdS/CFT correspondence and the Jarzynski identity as presented in Ref. \refcite{jarzads}. Next, in section 6 we review the proposal presented in Ref. \refcite{dmturb} regarding the emergent symmetries in fully developed turbulence and the corresponding non-equilibrium stationary states. Finally,
in section 7 we conclude our discussion by presenting a much larger picture regarding the relationship between non-perturbative quantum gravity and non-equilibrium statistical physics. 
According to this picture, the AdS/CFT duality is just an example of a more general connection between quantum gravity and non-equilibrium physics. 
In the appendix of the review we discuss a new kind of complementarity between thermodynamics and statistical physics which should be important in the context of black hole complementarity.
Our aim in this review is to emphasize the natural relationship between non-perturbative quantum gravity, in the context of string theory, and non-equilibrium physics.
This relationship can be understood as a natural generalization of the well-known Wilsonian relation between local quantum field theory and equilibrium statistical physics of critical phenomena.
We expect that this new relation between quantum gravity and non-equilibrium physics will be as illuminating and insightful as the well-known Wilsonian paradigm.

\section{From Quantum Gravity to Non-equilibrium physics}

We start our discussion by a brief review of the canonical thinking about gravitational thermodynamics and the reason for
the deep connection between quantum gravity and non-equilibrium physics.
(For the comprehensive treatment of this subtle issue see Ref. \refcite{freidel}.)
The basic intuition of black hole thermodynamics \cite{thermo,thermo1,thermo2} (derived from
Einstein's equations) can be turned on its head
to claim that
the Einstein's equations can be understood thermodynamically \cite{ted,ted1,verlinde,verlinde1}.
The first law is naively written as
\be 
dE = T dS 
\ee
where according to the second law $dS \ge 0$. Here $T = \frac{\hbar a}{2 \pi c}$ is the Hawking-Unruh temperature and $a$ is the local acceleration, whereas the entropy is given by the holographic Bekenstein-Hawking formula 
\be 
S= \frac{A c^3}{4 G_N \hbar} ~.
\ee

Notice that the gravitational entropy is not extensive, i.e. it does not scale with the volume of the system, but with the
boundary of the volume, i.e. the area $A$. (For the subtlety of the concept of entropy in the non-inertial frames see Ref. \refcite{mmr}.)
The claim is that Einstein's gravitational equations follow from these two laws of gravitational thermodynamics.
The factors of $\hbar$ cancel between the temperature and entropy
(temperature being perturbative in $\hbar$ and entropy being non-perturbative in $\hbar$)
in Einstein's gravitational equations.
Also the usual $8 \pi G_N$ factor from Einstein's equations
is now represented as $8 \pi G_N = 2\pi (4 G_N)$,
where the $2\pi$ factor (from the formula for $T$) is fixed by the relation
between the imaginary (or Euclidean) time $t_E$  in the exponential measure of the Euclidean path integral $\exp(-\frac{S_E}{\hbar})$ 
($S_E$ being the Euclidean action) and the
Boltzmann-Gibbs equilibrium exponential measure $\exp(- \beta H)$, $H$ being the Hamiltonian, i.e.
$\beta \hbar = t_E$,
the Euclidean time also being periodic with the period determined by the acceleration of the particle
$2 \pi c/a$.

Thus according to this intuition the Einstein equations essentially are the expression of the first law combined with the
second law $dS \ge 0$ which is implied by the focusing theorem (i.e. the attractive nature of gravity). \cite{ted,ted1}
The Raychaudhuri focusing theorem translates $dA$ into the space-time curvature $R_{ab}$\cite{mtw,hawkel}.
{\it Our first basic observation is that this reasoning implies an underlying non-equilibrium description!
The temperature is only locally defined and it is given by the local acceleration, or the local gravitational
field, as implied by the equivalence principle. Thus gravitational thermodynamics should be essentially
of the non-equilibrium character and thus quantum gravity, especially its locally holographic formulation,
should be essentially mapped onto non-equilibrium physics.}
(Note that some prescient statements about the relation between non-equilibrium physics
(especially the fluctuation-dissipation theorem and Onsager's relations) and
black hole thermodynamics can be found in Refs. \refcite{sciama,sciama1}.)

\section{Systems far from equilibrium}

\noindent
Next we turn to a brief review of the current research of systems far from equilibrium.
Understanding systems far from equilibrium is one of the most challenging problems in contemporary physics.
It is also one of the most important ones, due to the omnipresence of non-equilibrium processes in nature.
In virtually every discipline one observes an increasing focus on processes that take place far from equilibrium.
Notwithstanding this strong interest in non-equilibrium phenomena, we still lack a 
unifying theoretical framework for the
description of non-equilibrium interacting many-body systems. Still, some remarkable recent progress has been
achieved through the study of specific model systems with generic properties, using
the large toolbox of numerical and analytical methods provided by statistical physics \cite{Sch95,Muk00,Hen08,Odo08,book10,cmz}. 

This progress has been achieved in three different areas. Firstly, the study of relaxation processes
taking place in systems suddenly brought out of equilibrium has yielded an increased understanding
of physical aging in non-frustrated systems, especially in cases where the system is characterized
by a single, algebraically growing length scale \cite{book10}. Secondly, the discovery of various work and fluctuation theorems
has provided very general statements valid for large classes of systems \cite{Eva93,Jar97,Cro99,Sei12}. Thirdly, the in-depth investigation of steady-state
properties, especially in one-dimensional transport models \cite{cmz}, has yielded a better understanding
of the differences between equilibrium and non-equilibrium steady states.

In the following we briefly discuss some of these recent results, as this will allow us to put 
into a broader context the research summarized in this review.

At first look a unifying description of processes far from equilibrium seems to be hopeless, due to the apparently erratic and
history-dependent properties in that regime. It is therefore very remarkable that
the out-of-equilibrium properties of many systems with slow dynamics can be organized in terms of a simple scaling picture.
Glasses are one example of such systems. In many cases they are made by rapidly cooling
({\em quenching}) a molten liquid to below some characteristic temperature threshold.
If this cooling happens sufficiently fast, normal crystallization no longer takes place
and the material remains in some non-equilibrium state.
Studying the mechanical properties of polymeric glasses quenched
to low temperatures,
Struick \cite{Stru78} observed that the time-dependent creep curves of the mechanical response
can all be mapped onto a single master curve. This master curve turned out to be the same
for very different materials, pointing to a universal behavior far from equilibrium.

These experiments, which were the first systematic studies of
aging in out-of-equilibrium systems, revealed the following important features of
physical aging:
(i) slow (i.e. non-exponential) dynamics, (ii) breaking of time-translation invariance, and
(iii) dynamical scaling \cite{book10}.

In recent years remarkable progress has been achieved in the study of the out-of-equilibrium
dynamical behavior, and especially of aging, in cases where the system is characterized by a
single dynamical length $L$ that grows as a power-law of time $t$: $L(t) \sim t^{1/z}$, where
$z$ is the dynamical exponent. This simple growth law is encountered in many systems, ranging
from ferromagnets undergoing phase-ordering \cite{Bra94} to reaction-diffusion systems \cite{Hen07}.
The theoretical description of aging phenomena in these systems starts from the observation
that dynamical correlation and response functions
transform in a specific way under the dynamical scale transformation $t \longrightarrow b^z t$,
$\vec{r}  \longrightarrow b  \vec{r}$, where $b$ is a scale factor.
Of special interest are the two-time correlation and response functions, $C(t,s,\vec{r})$ and $R(t,s,\vec{r})$,
that are formally defined through the equations (here we use the language of magnetic systems,
but this can of course be generalized easily)
$C(t,s,\vec{r}) = \langle \phi(t,\vec{r})  \phi(s,\vec{0}) \rangle$ and
$R(t,s,\vec{r}) =  \delta\langle \phi(t,\vec{r})\rangle/\delta h(s,\vec{0}) |_{h=0}.$
Here $\phi(t,\vec{r})$ is the space-time-dependent order-parameter,
$h(s,\vec{r})$ is
the conjugate magnetic field, and $s$ and $t > s$ are two different times called waiting and
observation time. In the special case $\vec{r} = 0$, these quantities yield the autocorrelation function $C(t,s)$ and
the autoresponse function $R(t,s)$.
Due to dynamical scaling, these quantities display the following simple scaling form
in the dynamical scaling or aging regime \cite{Cug03,Cal05} where all the times involved are
large compared to any microscopic time scale:
$C(t,s) = s^{-b} f_C(t/s)$ and $R(t,s) = s^{-1-a} f_R(t/s)$,
where $f_C$ and $f_R$ are dynamical scaling functions, whereas $a$ and $b$ are non-equilibrium exponents.
The dynamical scaling functions of two-time quantities and the associated non-equilibrium exponents are
often found to be universal and to depend only on some global features
of the system under investigation.

The presence of dynamical scaling yields some hope that symmetry principles could be used successfully
for the description of this type of non-equilibrium processes, along lines similar to those that yielded
the successful use of conformal invariance in equilibrium critical phenomena. 

The theory of local scale invariance \cite{book10,Hen01,Hen07a} presents a novel approach
for the analytical treatment of stochastic systems. Starting from the stochastic Langevin
equation for some quantity of interest (as for example the order parameter in magnetic
systems), it proposes to split this equation into a {\it deterministic} part and a {\it noise}
part. A crucial observation is thereby that the deterministic part admits much richer
space-time symmetries than mere dynamical scale invariance.
Exploiting these non-trivial symmetries, all averages within the initial
noisy theory can be reduced exactly to averages within the deterministic, noiseless theory.
This is a very general result as it depends only on the (generalized)
Galilean invariance of the deterministic part.

Interestingly, the space-time symmetries of the noiseless theory strongly constrain the possible
forms of $n$-point functions. Especially for two-point functions, as the two-time response
and correlation functions, the scaling functions are completely fixed due to these symmetries.
This allows the model-independent
determination of universal scaling functions of response and correlation functions in aging
systems. The scaling functions obtained from this approach have been found to be correct
in exactly solvable models and to describe faithfully over many time decades the numerically determined scaling functions
of more complex systems \cite{book10}.

A second research thrust focuses on generic fluctuation properties of systems driven out of a steady state
\cite{Jar97,Cro99,Cro00,Hat01}. The progress in this field is closely related to the discovery of various fluctuation and work theorems
that provide generic statements applicable to large classes of systems. One well known example is the Crooks relation
where a system in equilibrium is driven out of equilibrium following a specific protocol. Repeating this process many times
allows to compute the probability distribution $P_F(W_d)$ of the dissipative work $W_d$. Using the time-reversed process one can
also determine the probability distribution for this process, $P_R$. Comparing these two distributions yields the simple
relation 
\begin{equation}
P_F(\beta W_d)/P_R(-\beta W_d) = e^{\beta W_d}~,
\end{equation}
where $\beta = 1/(k_BT)$. For the special case of equilibrium initial and
final states one can also relate the free energy difference to the following average over all processes
leading from one state to the other,
$\left< e^{- \beta W} \right> = e^{- \beta \Delta G}$,
with $W$ being the work done on the system, and $G$ the free energy. This is the celebrated Jarzynski work theorem.

The Jarzynski and Crooks relations have been generalized to many different situations, and detailed and integral fluctuation theorems
have been proposed for a range of quantities, from driving entropy production \cite{Hat01,Avr11a} to adiabatic and non-adiabatic
trajectory entropies \cite{Esp10}. 

Finally, many efforts have also been devoted to gain a better understanding and characterize more completely non-equilibrium steady states.
A well known example for a non-equilibrium steady state is of course the steady state realized in fully developed turbulence \cite{Kra80}.
Even though turbulence is far from being well understood, remarkable recent progress has been achieved in the characterization
of the steady states in simpler model systems. Thus for one-dimensional transport models, which to a large extend can be solved analytically,
a very detailed understanding of non-equilibrium steady states has been achieved \cite{cmz}. One of the hallmark of non-equilibrium steady states
is entropy production. While entropy
production is a system-dependent quantity, some remarkable universal properties have been discovered. Thus, the probability distribution
of the total entropy production satisfies a detailed fluctuation theorem in large classes of systems 
(see, e.g., Refs. \refcite{Eva93,Gal95,Kur98,Leb99,Sei05}), and a kink
appears in its large deviation function (and in that of related
currents) at zero entropy production \cite{Sei12,Meh08,Dor11,Avr11b}.

In the following three sections we turn to the unexpected relation between string theory and these research thrusts:
the aging phenomena, the Jarzynski identity and the steady states in fully developed turbulence.

\section{Aging Phenomena and String Theory}     %

In this section we review the connection between aging phenomena and string theory. The presentation is
based on our previous work. \cite{age1, age2, age3}.
Let us start by stating in more detail some of the known specifics about the dynamical scaling and scale invariance in
dynamical systems with aging \cite{Cug03,Hen07a}.
The general set-up for the study of aging behavior is as follows:
One considers a coarse-grained order parameter $\phi(t,\vec{r})$, conjugate to a generalized
field $h(t, \vec{r})$, which is usually assumed to be fully disordered
at $t=0$, i.e. $\langle \phi(0, \vec{r} )\rangle = 0$. For a magnetic system, the
order parameter is of course the magnetization whereas the conjugate field is an external magnetic field.
In the following we consider the case where the order parameter is
not conserved by the dynamics (model $A$ dynamics). As already pointed out in the previous section, one studies the 
scaling behavior of two-time correlation functions (here and in the following we assume that
spatial translation invariance holds)
\be
C(t, s) = \langle \phi(t, \vec{r}) \phi(s, \vec{r}) \rangle \sim s^{-b} f_C(t/s)
\ee
as well as of two-time response functions
\be \label{scale1}
R(t,s) = \frac{\delta \langle \phi(t, \vec{r} ) \rangle}{ \delta h(t, \vec{r})} \sim s^{-1-a} f_R(t/s)
\ee
where $a$ and $b$ are non-equilibrium exponents whereas $f_C$ and $f_R$ are scaling functions.
This scaling behavior is expected in the aging regime, defined by both $t$ and $s$ as well as
$t-s$ being much larger than the characteristic microscopic time scale.
Note also that this scaling assumes a single characteristic length scale $L$ which
scales with time $t$ as
\be
L(t) \sim t^{1/z}
\ee
where $z$ is the dynamical exponent.
For large values of the argument one expects
\be \label{scale2}
f_C(x) \sim x^{-\lambda_C/z}, \quad f_R(x) \sim x^{-\lambda_R/z}
\ee
with new non-equilibrium exponents $\lambda_C$ and $\lambda_R$.
The aging behavior just summarized is called simple or full aging and has been observed in many
exactly solvable models, in numerical simulations of more complex models as well as in actual experiments.

Given the success of conformal invariance in equilibrium critical phenomena,
it is natural to ask whether scaling functions and the values of non-equilibrium exponents might
be deduced from symmetry principles by invoking generalized dynamical
scaling with a space-time dependent scale factor $b = b(t, \vec{r})$.
This program has been reviewed in Ref. \refcite{Hen07a} and here we concentrate on the
specific case when $L(t) \sim t^{1/2}$, i.e.
$z=2$. This is a very important case as it encompasses systems undergoing phase-ordering
with non-conserved dynamics \cite{Bra94}. The generalization for $z \neq 2$ has been discussed in Ref. \refcite{Hen07a}.

The theoretical description of aging systems with a dynamical exponent $z=2$ starts from a stochastic
Langevin equation which for a non-conserved order parameter reads:
\be \label{langevin}
2 M \partial_t \phi = \nabla^2 \phi - \frac{\delta V[\phi]}{\delta \phi} + \eta
\ee
where $V$ is a Ginzburg-Landau potential and $\eta$ is a Gaussian white noise that arises due to the
contact with a heat bath \cite{footnote1}.
The theory of local scale invariance then permits to show \cite{Hen07a, Picone:2004id}
that under rather general conditions all averages (i.e., correlation and response functions) of the noisy theory 
(\ref{langevin}) can be reduced {\it exactly} to averages of the corresponding deterministic, noiseless theory.

For the $z=2$ case the relevant symmetry structure is the Schr\"{o}dinger group \cite{schrgrp,schrgrp1,schrgrp2}.
It is well known that the free diffusion equation 
\be
2 M \partial_t \phi = \nabla^2 \phi
\ee
is invariant under the Schr\"{o}dinger group (the free diffusion equation being essentially
equivalent to the free Schr\"{o}dinger equation).
The Schr\"{o}dinger group is defined through the space-time transformations
\be
t \to t' = \frac{\alpha t + \beta}{\gamma t+\delta}, \quad \vec{r} \to \vec{r'} = 
\frac{{\bf R} \vec{r} + \vec{v} t + \vec{a}}{\gamma t+\delta}
\ee
where $\alpha, \beta, \gamma, \delta$ and $\vec{v}, \vec{a}$ are real parameters and 
$\alpha \delta- \beta \gamma = 1$, whereas
$\bf{R}$ denotes a rotation matrix in $d$ spatial dimensions.

The Schr\"odinger group is an action on space and time coordinates extending the usual
Galilean symmetries to include anisotropic scaling $\vec x\to\lambda \vec x, t\to
\lambda^z t$. Throughout this section, we will only discuss the non-relativistic case,
$z=2$. These actions close on a larger group generated by temporal translations $H$,
spatial translations $P_{i}$, Galilean boosts $K_{i}$, rotations $M_{ij}$, dilatations $D$
and the special `conformal' transformation $C$.

The commutation relations satisfied by these generators, aside from the obvious
commutators of rotation and translations and those that vanish, read\cite{age2}

\begin{align}\label{eq:Schrdalg}
\left[D,K_{i}\right] &= K_{i},& \left[D,P_{i}\right] &= -P_{i},& \left[P_{i},K_{j}\right] &= \delta_{ij}N\\
 \left[D,C\right] &= 2C,&  \left[C,P_{i}\right] &= -K_{i},&  \left[H, C\right] &= D\\
 \left[H,K_{i}\right] &= P_{i},& \left[D,H\right] &= -2H\label{eq:Schrdalg3}
\end{align}
We note that $\{H,D,C\}$ generate an $SL(2,\mathbb{R})$ subalgebra, with $\{P_i,K_i\}$ forming an $SL(2,\mathbb{R})$ doublet, and the generator $N$ is central.
We will refer to the full algebra as $Schr_d$, where $d$ is the spatial dimension. This algebra made its first appearance long ago, for example, as the invariance group of the Schr\"odinger equation with zero potential.

If one considers field theories with this symmetry, one finds, in complete analogy with
the conformal field theory bootstrap, \cite{Belavin:1984vu} that correlation functions are
of a restricted form \cite{Hen07a}. For example, for operators that are scalars under rotations,
the two-point function is essentially given by 
\be
\langle {\cal O}_1(t_1, \vec{x}_1){\cal O}_2(t_2, \vec{x}_2) \rangle= \delta_{\Delta_1, \Delta_2} \delta_{n_1+n_2,0}
(t_{1}-t_{2})^{-\Delta_1} \exp\left(i\frac{n_2}{2} \frac{(\vec{x}_{1}-\vec x_{2})^{\,2}}{t_{1}-t_{2}}\right)\, .
\ee
In the holographic context, this result
was obtained using real-time methods \cite{Skenderis:2008dh,Skenderis:2008dg}  in Ref. \refcite{Leigh:2009eb}, and using Euclidean methods in Refs. \refcite{Fuertes:2009ex,Volovich:2009yh}. Similarly, higher point functions are of a constrained form as well.

In holography, one makes use of a space-time possessing $Schr_d$ as its algebra of
isometries. Such a space-time may be taken to have
metric \cite{Son:2008ye,Balasubramanian:2008dm}
\begin{equation}\label{Son-McGreevy}
ds^{2} = \frac{L^{2}}{z^{2}}\left(dz^{2} - \frac{\beta^{2}}{z^{2}}dt^{2} - 2dtd\xi + d\vec{x}^{2}\right),
\end{equation}
where $\vec{x} = (x_{1},\ldots, x_{d})$ are the spatial coordinates of the dual field theory and
$z$ is the holographic direction (with the ``boundary" located at $z = 0$). The parameters
$\beta$ and $L$ are length scales, and all the coordinates have units of length.

The space-time \eqref{Son-McGreevy} has Killing vectors
\beqn\label{general Killing vector nu one}
M_{ij}&=& x_{j}\partial_{i} - x_{i}\partial_{j}\\
P_{i} &=& \partial_{i},\ \ \ H= \partial_{t}, \ \ \ \ N= \partial_{\xi}\\
C &=& zt\,\partial_{z} + t^{2}\,\partial_{t} + \frac{1}{2}\left(z^{2} + \vec x^{2}\right)\partial_{\xi} + tx^{i}\,\partial_{i}\\
D&=& z\,\partial_{z} + 2t\,\partial_{t} + x^{i}\partial_{i}\\
K_{i} &=&  t\,\partial_{i} + x^{i}\partial_{\xi} 
\eeqn
providing a representation of the $Schr_d$ algebra (\ref{eq:Schrdalg}--\ref{eq:Schrdalg3}) acting on bulk (scalar) fields. Since $N$ is central, such fields can be taken to be equivariant with respect to $N$, i.e., their $\xi$-dependence can be taken to be of the form $e^{in\xi}$ for fixed\footnote{$\xi$ is often taken to be compact, so that the spectrum of operators in the dual theory is discrete. This is not without problems in the bulk, as $\pa_\xi$ is null in the geometry (\ref{Son-McGreevy}).} $n\in\mathbb{R}$.  In gauge/gravity
duality, the asymptotic ($z\to 0$) values of fields propagating in this geometry act as sources for
operators in the dual field theory. The $Schr_d$ algebra acts on those operators in a way
that can be deduced from the field asymptotics \cite{Leigh:2009eb,Leigh:2009ck}. We thus
get another distinct representation of $Schr_d$ that acts on (scalar) operators of the dual field theory
\beqn
M_{ij}&=& x_{j}\partial_{i} - x_{i}\partial_{j}\\
P_{i} &=& \partial_{i},\ \ \ H= \partial_{t}, \ \ \ \ N= \partial_{\xi}\\
D&=& 2t\,\partial_{t} + x\partial_{x} + \Delta\, \mathbb{1}\\
C &=&  t^{2}\,\partial_{t}  + tx\,\partial_{x} + \frac{x^{2}}{2}\,\partial_{\xi}+ \Delta t\, \mathbb{1}\\
K_i&=&  t\,\partial_{i} + x_i\partial_{\xi} 
\eeqn
where $\Delta$ is the scaling dimension; for equivariant fields, $\pa_\xi$ evaluates to $in$.

\subsection{The Aging Algebra and Correlation Functions}\label{sec:aging}  %

The aging algebra, which we will denote as $Age_d$, is obtained by discarding the time
translation generator $H$ from the $Schr_d$ algebra. Indeed, the form of the algebra is
such that it makes sense to do so. This is the simplest possible notion of time-dependent
dynamics, and it is considered as a rather special form of non-equilibrium physics. 
We consider the problem of constructing an appropriate space-time geometry possessing $Age_d$ as its isometry algebra, and then compute some simple correlation functions, following
Ref. \refcite{age2}.

Since $H$ has been discarded, such correlation functions are not generically time-translation invariant. To see what sort of time-dependence to expect, let us consider a construction that often appears in the literature.
Consider a diffusive system with (white) noise
$\eta$ and a time-dependent potential $v(t)$ governed by a wave-function $\phi$ satisfying\footnote{The
diffusion equation should be regarded as a Wick rotated version of a Schr\"odinger-like
equation (or equivalently, a Schr\"odinger-like equation is obtained by considering $M$ to
be imaginary). In following sections, we will always work in Lorentzian signature. One does not expect that such Wick rotations are innocuous in general.}
\be
\label{dyneqn}
2 M \partial_t \phi = \nabla^2 \phi - \frac{\delta V}{\delta \phi} - v(t) \phi + \eta.
\ee
As pointed out in Ref. \refcite{book10}, correlation functions in this system can be studied by examining deterministic
dynamics governed by the aging group. Note that the `gauge transformation'
\be
\label{GaugeTransformation}
\phi \to \phi \exp\left(-\frac{1}{2M} \int^t v(\tau) d \tau\right),
\ee
removes the time dependent potential term from equation (\ref{dyneqn}).
This means that this sort of time-dependence can be mapped to a system governed by   the Schr\"odinger group. A simple physical model
for this type of time dependence is the out of equilibrium decay of a system towards equilibrium, following some sort of quench.
In the special case
\be
v(t) \sim 2M Kt^{-1},
\ee
one finds that the wavefunctions are related via a scaling function to the wavefunctions of the Schr\"odinger problem
\be
\label{aging-schrodinger}
\phi_{Age} = t^K \phi_{Schr}.
\ee
%
Thus, the local scale-invariance of aging systems is largely determined by studying first
the Schr\"odinger fields. In particular, the correlators of operators in an $Age_d$-invariant theory can be expressed in terms of  the correlators of operators in a $Schr_d$-invariant theory. Schematically, for the two-point function of a scalar operator $\mathcal{O}$, the result we want to reproduce is \cite{book10,Picone:2004id}

\begin{equation}\label{schematic expected form of correlator}
\langle \mathcal{O}(t_{1})\mathcal{O}(t_{2})\rangle_{Age} \sim \left(\frac{t_{1}}{t_{2}}\right)^{\#}\langle \mathcal{O}(t_{1})\mathcal{O}(t_{2})\rangle_{Schr},
\end{equation}

\noindent where $\#$ is a constant which characterizes the breaking of time-translation invariance. We will present the details of this relationship below.

Above, we gave two representations of $Schr_d$, one acting on (scalar) operators of a field theory,
and one acting on the holographic bulk (scalar) fields.
We now consider removing $H$ from the algebra, and ask how the representation of the
remaining generators might be modified. We will consider first the representation on operators, as that is what appears 
in the literature. Let us assume that the  time and spatial coordinates
$t$ and $\vec x$ and the non-relativistic mass (equivalently the coordinate $\xi$) retain their
standard meaning. Consequently, we take the representation of $M_{ij},P_i,N,K_i$ and $D$ to be unchanged from that of $Schr_d$.
However, it is possible that the form of the generator $C$ could be modified consistent with the $Age_d$ algebra. Suppose then that we write
$C_{A} = C + \delta C$, where $C_{A}$ is the
representation of $C$ in the aging algebra. In order for $C_{A}$ to satisfy the
commutators of the $Age_d$ algebra (which are unchanged from $Schr_d$) we need

\begin{equation}
\left[P_i,\delta C\right]=0, \qquad \left[N,\delta C\right]=0, \qquad \left[K_i,\delta C\right]=0, \qquad \left[D,\delta C\right] = 2\delta C.
\end{equation}
The first three commutators are easily seen to imply that $\delta C$ can only be of the form
$\delta C = g_{1}(t)\mathbb{1} + g_{2}(t)\partial_{\xi}$. The fourth commutator then fixes $g_{i}(t)$ ($i=1,2$):
\begin{equation}
\left[D,\delta C\right] = 2\delta C \qquad \Rightarrow \qquad t\partial_{t}g_{i}(t) = g_{i}(t) \qquad \Rightarrow \qquad g_{i}(t) = K_{i}t,
\end{equation}
and we then conclude \cite{Picone:2004id} that the most general form is
\begin{equation}\label{C in Aging}
C_{A} =  t^{2}\,\partial_{t}  + tx\,\partial_{x} + \frac{x^{2}}{2}\,\partial_{\xi} + (\Delta + K_1) t\, \mathbb{1} + K_2t\,\partial_{\xi},
\end{equation}
where $K_1$ and $K_2$ are constants. Note that in the undeformed case $K_1=K_2=0$, and that when evaluated on equivariant fields, $K_2$ is accompanied by the eigenvalue of $\partial_\xi$ which is imaginary. Thus, $\delta C = Kt\,\mathbb{1}$, where $K=K_1+inK_2$ is naturally a complex number. The physical significance of the real and imaginary parts will be discussed more fully below, but for now we note that we expect $K$ to make an appearance in time non-translation invariant features of correlation functions. 

$K$ has been described in the literature \cite{Picone:2004id} as a `quantum number' labeling representations. As pointed out in Ref. \refcite{age2}, we believe that it is better to think of this parameter as a property of the {\it algebra} instead; indeed, later we will see $K$ emerge in a holographic setup as a parameter appearing in the metric, rather than being associated with any particular field. We further note that if we were to demand $\left[\pa_t,C_{A}\right] =D$, we  would find that $K=0$, thus recovering the full Schr\"odinger algebra in its
standard representation.


\subsection{A Geometric Realization of the Aging Group }\label{Sec:Metrics}   %

Next, following Ref. \refcite{age2}, we will explore how we might implement $Age_d$ as an isometry algebra.
There are a variety of possible constructions that present themselves. First, we might consider the breaking
of time translation invariance to be associated with the introduction of a `temporal defect'. In relativistic gauge/gravity duality, there is a way to introduce spatial defects \cite{DeWolfe:2001pq,Aharony:2003qf} by placing a D-brane along an $AdS_{d-1}$ slice of $AdS_d$. Such a brane intersects the boundary along a co-dimension one curve, which can be coordinatized as $x=0$. The choice of slicing preserves as much symmetry as possible: clearly $P_x$ is broken along with $K_x$ and $M_{xj}$,\footnote{Here we refer to the generators of the $AdS_d$ isometry, $SO(d,2)$.} leaving unbroken $SO(d-1,2)$. Because of Lorentz invariance, presumably such a construction works for temporal defects in the relativistic case (see Ref. \refcite{Bak:2007qw} for related work on time-dependent holography). Such a construction would involve placing an S-brane along a suitable slice.

Similarly, spatial defects \cite{Karch:2009rj} can be introduced in non-relativistic holography in much the same way. One can imagine placing a D-brane along a $Schr_{d-1}$ slice of the $Schr_d$ space-time. Here, $P_x$, $K_x$ and $M_{xj}$ would be broken. It is fairly obvious though that in this case, a temporal defect cannot be constructed in this way. The basic reason is that time is much different than space in a non-relativistic theory, and at the algebraic level, we seek to break $H$ only. We will see signs of this sort of difficulty below.

The key feature of such metrics is that $Age_d$ invariance allows the metric components to depend on the invariant combination $T=\beta t/z^2$. This is scale invariant (and dimensionless) but clearly transforms under time translations. A fairly general Ansatz for the metric is then of the form\cite{age2}
\begin{align}\label{metric ansatz}
ds^{2} &= \frac{L^{2}}{z^2}\left[f_{zz}\left(T\right)\,dz^2 + \frac{2\beta}{z}f_{zt}\left(T\right)\,dzdt - \frac{\beta^{2}}{z^{2}}f_{tt}\left(T\right)\,dt^{2} + f\left(T\right)\left(-2dtd\xi  + d\vec x^{2}\right)\right]\, .
\end{align}
As a regularity requirement, the functions $f_{zz}$ and $f$ are
chosen such that $\det g =-L^{8}f_{zz}f^{3}/z^{8}$ is non-zero everywhere other than
$z=\infty$. It is clear that the Ansatz \eqref{metric ansatz} already admits $\{M_{ij},P_i,K_i,D,N\}$ as
isometries and the generators are in the same form as in the Schr\"odinger space-time, (\ref{general Killing vector nu one}). If the metric functions $f_{zz}(T)$ and
$f(T)$ are independent, one concludes that there are no further isometries.

Although in this construction the full $Schr_d$ algebra is present locally, we clearly need to be careful near $t=0$. If we interpret the blowing up of the components of $H$ as an indication that we lose $H$ as a generator, we obtain precisely what we want in the dual field theory. It should be emphasized again however that this is a coordinate singularity; for example, the norm of the vector field $H$ is well-behaved everywhere in the metric\begin{align}\label{the potential age metric}
ds^{2} &= \frac{L^{2}}{z^2}\left[dz^2 + \frac{2\alpha\beta}{z}\,dzdt - \frac{\beta^{2}}{z^{2}}\left(1 + \frac{\alpha}{T}\right)\,dt^{2} -2dtd\xi  + d\vec{x}^{2}\right].
\end{align}
We note also that this metric breaks a discrete symmetry enjoyed by the $Schr_d$ metric, namely `$CT$', $t \to -t, \xi \to -\xi$. 
The following change of coordinates would take this metric back to the standard Schr\"odinger form,
\begin{equation}
\label{xi-log-trans}
\xi' = \xi + \frac{\alpha\beta}{2}\ln T = \xi + \frac{\alpha\beta}{2}\ln\left(\frac{\beta t}{z^{2}}\right)\, .
\end{equation}
The multi-valuedness of the logarithm in the complex $t$ plane will play an important role in the physical interpretation of the point $t=0$ and in the calculation of correlation functions that we present below.

\subsection{Correlation Functions}\label{sec:correlators} %

Now we review the correlations functions in the background of the aging geometry as presented in Ref. \refcite{age2}.
Consider a scalar field on the geometry (\ref{the potential age metric}). We look for solutions to the scalar wave equation for the above aging geometry that are of the form
\begin{equation}\label{new separable ansatz}
\phi(z,t,\xi, x) = e^{Q(T)}e^{in\xi}\phi_{S}(z,t,x; n) = e^{Q(T)}e^{in\xi}\int \frac{d^dp}{(2\pi)^d} \frac{d\omega}{2\pi}\, e^{i\vec p\cdot\vec x - i\omega t}\phi_{S}(z,\omega, p; n),
\end{equation}

\noindent where $Q(T)$ is some function  and $\phi_{S}$ denotes a solution of the wave equation on the standard Schr\"odinger background \eqref{Son-McGreevy} (which is translationally invariant). If such a form exists, then the appearance of time translation non-invariance in the scalar is entirely in  the scale-invariant prefactor $\exp(Q(T))$. We note that given the fact that the form of the dilatation operator is the same for both $Age_d$ and $Schr_d$, the corresponding fields have the same conformal dimension, even though there is $z$-dependence in the prefactor, precisely because $T$ is dilatation invariant. We note also that the Ansatz for the scalar fields is reminiscent of the expected form (\ref{aging-schrodinger}).

Indeed, for the geometry (\ref{the potential age metric}), the scalar Laplacian is such that solutions are of the form (\ref{new separable ansatz}), with
\begin{equation}\label{Q of T}
Q(T)= \frac{in\beta}{2}\int^T dT'\left[f_{tt}(T') - 1\right].
\end{equation}
 For the Schr\"odinger field, in the asymptotic region we have
\beq
\phi_S(z\to 0,t,\vec x;n)\sim z^{\Delta_-} \phi^{(0)}_S(t,\vec x;n)+...
\eeq
and thus the aging field behaves as
\beq
\phi(z\to 0,t,\vec x;n)\sim z^{\Delta_-} \left(\frac{\beta t}{z^2}\right)^{-in\beta\alpha/2}\phi^{(0)}_S(t,\vec x;n)+...
\eeq
As we mentioned, although there is an unusual factor of $z$ present here, because the generator $D$ is unmodified, the scaling dimension of $\phi$ is the same as $\phi_S$. As we will see below, the extra factor of $z$ should be thought of as a `wavefunction renormalization' factor that should be absorbed into the definition of  the dual operator --- it is included in the source for Age fields. 

Now, one might assume since the prefactor is common to both the source and the vacuum expectation value, that it cancels out in the evaluation of the Green function. We will show below that this is naive  --- it is related to depending too much on Euclidean methods. When one carefully examines the real-time correlation functions, one finds that the prefactor makes its presence felt.
Note also that the prefactor is generally complex. If we confine ourselves to $\alpha\in\mathbb{R}$, which one normally would do in real geometry, then formally the prefactor is a phase. However, $Q(T)$ is generally a complex function, and in our case, it is a multi-valued function about $t=0$. It is then not too much of a stretch to go all the way to a complexified geometry, in the sense of taking $\alpha\in\mathbb{C}$. As we have discussed above, the real and imaginary parts of $\alpha$ (or $K$ in the language of Section \ref{sec:aging}) have a direct physical interpretation, and we will encounter precisely that in the $Age_d$-invariant correlation functions that we study below.

The Schr\"odinger theory is renormalizable \cite{Leigh:2009eb}; the on-shell action can be made finite by the inclusion of a number of {\it Schr\"odinger invariant} counterterms. One can show that the boundary renormalization of the Age theory follows that of Schr\"odinger closely. Indeed, the bulk action
\beq
S = -\frac{1}{2} \int d^{d+3}x \sqrt{-g} \left( g^{\mu\nu}\pa_\mu \bar\phi\pa_\nu \phi +
m_0^2/L^2 |\phi|^2 \right)\label{action}
\eeq
reduces on-shell to the boundary term
\begin{align}
S_{os} = \frac{1}{2}\int_\epsilon d^{d+1}x d\xi\sqrt{|\gamma|}
\ \bar\phi {\bf n}\cdot\pa\phi
\end{align}
Here we take  the `boundary' to be a constant $z$ slice, with normal $n=\frac{L}{z}dz$;  the corresponding normal vector is ${\bf n}=\frac{1}{L}(z\pa_z+\alpha\beta\pa_\xi)$ when we use $\ftt=1+\alpha/T$. Since the on-shell solutions are of the form $\phi=e^{Q(T)}\phi_{Schr}$, we get
\begin{align}\label{eq:onshellactionAge}
S_{os} = \frac{1}{2L}\int_\epsilon d^{d+1}x d\xi\sqrt{|\gamma|}
\ e^{Q(T)+\overline{Q(T)}}\bar\phi_{Schr} z\pa_z\phi_{Schr}
\end{align}
with $\gamma$ the induced spatial metric. This is precisely of the same form as in the Schr\"odinger case, with the inclusion of the prefactors appropriate to Age fields. We see again that one must be careful in this case in interpreting powers of $z$ --- since the Age fields are of dimension $\Delta$, the scale invariant $z$-dependent prefactors must be associated with the normalization of operators. They are not to be canceled by the addition of counterterms.

As in Ref. \refcite{age2}, we will focus on the calculation of the two-point functions of scalar operators dual to $\phi$. As we stated above, because of the time-dependence of the metric, it is dangerous to attempt to employ Euclidean continuation, and thus, as in Ref. \refcite{age2} we will consider
the real time correlators very carefully using the Skenderis-Van Rees method \cite{Skenderis:2008dh,Skenderis:2008dg},  following \refcite{Leigh:2009eb} closely.

\subsubsection{Review of Schr\"odinger calculations}

\begin{figure}[ht]
	\centering
	\includegraphics[height=7cm]{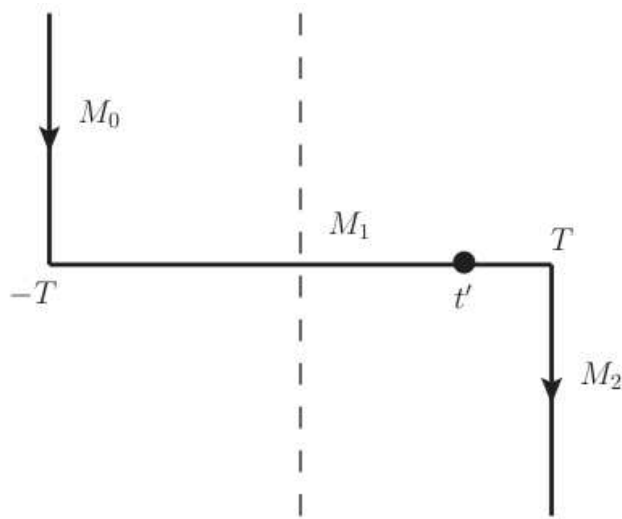}\caption{Contour in the complex $t$-plane corresponding to the time-ordered correlator.}\label{fig:Keldysh2}
\end{figure}

We now outline the results found in Ref. \refcite{Leigh:2009eb} for the case of the Schr\"odinger geometry. Generally, correlators are computed by constructing solutions along segments of a contour in the complex time plane, where the choice of contour determines the nature of the correlator considered. In the case of a time ordered correlator, the contour is as shown in Fig. \ref{fig:Keldysh2}.

A vertical segment corresponds to Euclidean signature, while a horizontal segment corresponds to Lorentzian signature. One computes the solutions in the presence of a $\delta$-function source on the real axis along each segment, and matches them at junctions.

\subsubsection{Schr\"odinger Solutions: Lorentzian}

In Lorentzian signature, we use the notation $q= \sqrt{q^2} = \sqrt{ \vec k^2-2\omega n}$. Solutions of the scalar wave equation on the Schr\"odinger space-time are given in terms of Bessel functions. For $q^2<0$, both the $K_\nu$ and $I_\nu$ solutions are regular everywhere, while for $q^2>0$, $I_\nu$ diverges for large $z$ and is discarded. Noting that $q$ has a branch point at $\omega=\vec k^2/2n$, we facilitate integration along real $\omega$ by properly deforming $q$ to $q_\epsilon=\sqrt{-2\omega n+\vec k^2-i\epsilon}$. With these comments, we arrive at  the general solution  in Lorentzian signature  \cite{Leigh:2009eb}
\beqn\label{general solu}
\phi_S(z,t,\vec x;\xi) = e^{in\xi} \int \frac{d\omega}{2\pi} \frac{d^d k}{(2\pi)^d}\
e^{-i\omega t + i \vec k\cdot  \vec x}  z^{\frac{d}{2}+1}
\left( A_s(\omega,\vec k) K_\nu(q_\epsilon z)
+ \theta(-q^2)B_s(\omega,\vec k) J_\nu(|q| z) \right)\, . 
\eeqn

\subsubsection{Schr\"odinger Solutions: Euclidean}

Next, we consider a similar analysis in Euclidean signature. To do so, we Wick rotate the Schr\"odinger metric to (more precisely, along $M_0$ and $M_2$ respectively, we write $t=\pm(-T+i\tau)$)
\beq
ds^2=\frac{L^2}{z^2}\left[ dz^2 +\frac{\beta^2}{z^2} dt^2-2id\tau d\xi+d\vec x^2\right]\, .
\eeq
Although this metric is complex, it is possible to trace carefully through the analysis. The general solution of the Schr\"odinger problem is
\beqn\label{E-general solu}
\phi_S(z,\tau,\vec x;\xi) &=& e^{in\xi} \int  \frac{d\omega_E}{2\pi} \frac{d^d k}{(2\pi)^d}\ e^{-i\omega_E \tau +
i \vec k\cdot  \vec x}  z^{\frac{d}{2}+1} A(\omega_E,\vec k) K_\nu(q_E z)\, ,
\eeqn
where now $q_E = \sqrt{q_E^2} = \sqrt{\vec k^2-i2\omega_E n}$. Note that in this case, the branch point is at
imaginary $\omega_E$, and so no $i\epsilon$ insertion is necessary. In writing this, we have assumed that $\tau\in(-\infty,\infty)$ and thus $\phi_S$ has no normalizable mode. One has to be careful with this. For example, if $\tau \in [0,\infty)$, we write $\omega_E = -i\omega$ for $\phi$ and
$\omega_E = i\omega$ for $\bar\phi$ and the following mode is
allowable
\beqn
\phi_S &\sim& e^{in\xi} e^{-\omega \tau + i \vec k\cdot \vec x}
z^{\frac{d}{2}+1} I_\nu(q_E z)
\eeqn
as long as $\omega > 0$ and $-2\omega n + \vec k^2 < 0$, or
equivalently $\omega > \vec k^2/2n$. For $\tau\in (-\infty,0]$, no such mode is present.

\subsubsection{Correlators (Schr\"odinger)}

The correlators are computed by recognizing the asymptotics as sources for corresponding operators
 \beq
 e^{iS^{bulk}_C[\bar\phi_S^{(0)},\phi_S^{(0)}]} = \langle
e^{i\int_C ( \hat {\cal O}^\dag\phi_S^{(0)} + \bar\phi_S^{(0)} \hat
{\cal O})}\rangle
\eeq
where the fields have asymptotic expansions
\beqn
\phi_S &=&  e^{in\xi}\Big\{ z^{\Delta_-} \left(\phi_{(0)}+ z^2 \phi_{(2)} + o(z^4)\right) + z^{\Delta_+} \left(v_{(0)} + z^2 v_{(2)} + o(z^4)\right)\Big\}\eeqn
with $\Delta_\pm=1+d/2\pm\nu$ and
\begin{align}
\phi_{(2m)} = \frac{1}{2m (2\Delta_+ - (d + 2) -
2m)}{\cal S}\phi_{(2m-2)}\, ,
\end{align}
where ${\cal S}=2in\partial_t+ \vec\nabla^2$ is the Schr\"odinger operator.
In Ref. \refcite{Leigh:2009eb} the  bulk to boundary propagator was derived
\beqn
K_{(n)}(t,\vec x,z; t')&=&\frac{2 z^{1+d/2}}{\Gamma(\nu)}  \int \frac{d\omega}{2\pi} \frac{d^d k}{(2\pi)^d}\ e^{-i\omega (t-t') + i \vec k\cdot \vec x} \left(\frac{q_\epsilon}{2}\right)^\nu
K_\nu(q_\epsilon z)\, ,
\eeqn
which determines the time ordered correlator
\beq\label{time-ordered}
 \langle T\Big{(}\hat
{\cal O}_{(n)}(\vec x,t) \hat {\cal O}_{(n)}^\dag(\vec
x',t')\Big{)}\rangle_{Schr.} =
\frac{1}{\pi^{d/2}\Gamma(\nu)}\left(\frac{n}{2i}\right)^{\Delta_+-1}
\frac{\theta(t-t')}{(t- t')^{\Delta_+}} e^{in\frac{(\vec x -
\vec x')^2 + i\epsilon}{2(t-t')} }.
 \eeq
The derivation of the final result involves a somewhat difficult contour integral.

\subsection{Aging Correlators}

Now, we turn our attention to the computation of correlators holographically in the Age geometry\cite{age2}. Because of the physical interpretation, that of a quench at $t=0$, we focus on correlators of operators inserted at times $t>0$. We have argued that the point $t=0$ is much like a horizon, and so one should in fact confine oneselves to the $t>0$ patch. As we have also discussed, the Age solutions have a discontinuity across $t=0$ given by $\phi(-\epsilon,\vec{x})=e^{i\pi n\beta\alpha/2}\phi(+\epsilon,\vec{x})$. To calculate the Age correlators, we propose that one should use the same Keldysh contour as above, fixing the solutions to the Age problem formally by requiring the continuity of the associated Schr\"odinger solutions. This prescription uniquely determines the two-point correlation function, and as we shall see, gives the expected form \eqref{schematic expected form of correlator}. We proceed to construct solutions on the segments $M_0, M_1$ and $M_2$.

\subsubsection{Age Solutions: Lorentzian}

With the same setup as for the Schr\"odinger problem above, we arrive at  the general solution  in Lorentzian signature
\beqn\label{general solu}
\phi(t,\xi,\vec x,z) = e^{Q(T)}e^{in\xi} \int \frac{d\omega}{2\pi} \frac{d^d k}{(2\pi)^d}\
e^{-i\omega t + i \vec k\cdot  \vec x}  z^{\frac{d}{2}+1}
\left( A_s(\omega,\vec k) K_\nu(q_\epsilon z)
+ \theta(-q^2)B_s(\omega,\vec k) J_\nu(|q| z) \right)
\eeqn

\subsubsection{Age Solutions: Euclidean}
\newcommand\fttE{f_{tt}(T_E)}

Next, we consider a similar analysis in Euclidean signature. To do so, we take complex time $t=\pm(-T+i\tau)$ (this is appropriate to $M_0$ and $M_2$ respectively) and thus replace the  Schr\"odinger metric by
\beq
ds^2=\frac{L^2}{z^2}\left[ dz^2 +\frac{\beta^2}{z^2}\fttE d\tau^2\mp 2id\tau d\xi\pm 2i\frac{\beta T_E}{z}(\fttE-1)d\tau dz+d\vec x^2\right]
\eeq
where $T_E=\pm\beta(-T+i\tau)/z^2$. The general solution of the Age problem is
\beqn\label{E-general solu}
\phi(\tau,\xi,\vec x,z) &=& e^{Q(T_E)}e^{in\xi} \int  \frac{d\omega_E}{2\pi} \frac{d^d k}{(2\pi)^d}\ e^{-i\omega_E \tau +i \vec k\cdot  \vec x}  z^{\frac{d}{2}+1} A(\omega_E,\vec k) K_\nu(q_E z)
\eeqn
where now $q_E = \sqrt{q_E^2} = \sqrt{\vec k^2-i2\omega_E n}$.

\subsubsection{Time-ordered Correlator}

As in the Schr\"odinger case, the solution along the $M_0$ component is zero. Thus we have to require $\phi_1(t_1=-T,\vec x,z)=0$. We then also conclude that there is no normalizable solution on $M_1$. Thus, we should have a unique solution.

We place a $\delta$-function source at $\vec x=0, t_1=\hat t_1>0$ on $M_1$. That is, we want
\beq
\phi_1(t_1,\vec x,z)\Big|_{z\to 0}=z^{\Delta_-}e^{in\xi}\delta(t_1-\hat t_1)\delta(\vec x)
\eeq
This requires the field to be of the form
\beq\label{phi1}
\phi_{1}(t_1,\vec x,z) =\frac{2}{\Gamma(\nu)}e^{Q(T_1)-Q(\hat T_1)}e^{in\xi}  z^{1+d/2-in\alpha\beta} \int  \frac{d\omega}{2\pi} \frac{d^d k}{(2\pi)^d}\ e^{-i\omega (t_1-\hat t_1) + i \vec k\cdot \vec x}
\left(\frac{q_\epsilon}{2}\right)^\nu K_\nu(q_\epsilon z).
\eeq

\noindent On $M_2$ we then have (where we match at $t=T$)

\begin{align}
\phi_{2}(\tau_{2},\vec{x},z) = \frac{2\pi i}{\Gamma(\nu)}& e^{Q(-i\beta(\tau_2+iT)/z^2)-Q(\hat T_1)}e^{in\xi}  z^{1+d/2-in\alpha\beta}\nonumber\\
&\times \int  \frac{d\omega}{2\pi} \frac{d^d k}{(2\pi)^d}\ e^{-\omega (\tau_2 + iT-i\hat t_1)  + i \vec k\cdot \vec x}
\theta(-q^2)\left(\frac{|q|}{2}\right)^\nu J_\nu(|q|z).
\end{align}

\noindent For $t_1>\hat t_1$, the correlator $K(t_1,\hat t_1)$ is essentially $\phi_1$ itself. Given the choice of $Q(T)$, we have
\beq
K_{age}(t_1,\hat t_1;\vec x)=\left(\frac{t_1}{\hat t_1}\right)^{in\alpha\beta /2} K_{Schr}(t_1,\hat t_1;\vec x).
\eeq
This result is the expected one --- it displays the time-translation non-invariant scaling form, with exponent given by $K=i\alpha n\beta/2$. We see that for real $\alpha$, this time dependence is a phase.\footnote{In the aging literature, many specific systems have been studied numerically for which no such phase is present. To be capable of seeing such a phase, one must at least have a complex order parameter. As an example, it is possible that such behavior could be seen in $p_{x}+ip_{y}$ superconductors.} It is only for $\alpha$ complex that the correlator corresponds to a relaxation process. For generic $\alpha$, the correlator `spirals in' towards the Sch\"odinger correlator at late times. 
Note that for this physical interpretation, one expects that ${\rm Im}\ \alpha>0$. This is related holographically to normalizability of the solutions. Indeed if one traces the solution back to $t=-T$, for ${\rm Im}\,\alpha <0$, the prefactor of the solution blows up as $T\to\infty$, but is innocuous for $\alpha$ in the upper half plane.

\subsection{Comments on the 3-Point functions ang aging geometry}

The above realization of aging does
not capture the most general aging dynamics and that what has been described above is
just a particular realization of the Schr\"odinger dynamics! This was discussed in Ref. \refcite{age3}.
In what follows we clearly distinguish between this special case and the most
general aging dynamics.

As in Ref. \refcite{age3}, for simplicity, let us consider a 1+1-dimensional theory with coordinates $t,r$. We use $\xi$ to denote the Fourier variable conjugate to the mass ${\cal M}$ of a certain primary operator. In the notation of Refs. \refcite{book10,hu}, the Schr\"odinger and Age algebras are respectively spanned by the generators $\{X_{-1}, X_0, X_1, M_0, Y_{\frac{1}{2}}, Y_{-\frac{1}{2}}\}$ and $\{ X_0, X_1, M_0, Y_{\frac{1}{2}}, Y_{-\frac{1}{2}}\}$, which obey the following commutation relations 
\bea
&&\!\!\!\!\![X_n, X_{n'}]=(n-n')X_{n+n'}, \;\;[X_n, Y_m]=(\tfrac {n}{2} -m)Y_{n+m}\nn\\
&&\!\!\!\!\![X_n, M_{n'}]=-n' M_{n+n'}, \;\; [Y_m, Y_{m'}]=(m-m')M_{m+m'}.
\eea
The most general realization of these generators presented in Ref. \refcite{age3} is
\bea
X_{-1} = -\partial_t+\frac{g(t)-\gamma}{t}+\frac{h(t)-\delta}{t} i \partial_\xi, \;
X_0 = -t\partial_t -\tfrac 12 r \partial_r -\tfrac {\Delta}2 + g(t)+h(t) i\partial_\xi, \nn\\
X_1 = -t^2 \partial_t - tr\partial_r - \Delta t + \tfrac i2 r^2 \partial_\xi + t(g(t)+\gamma)+t(h(t)+\delta)i\partial_\xi, \nn\\
Y_{-\frac{1}{2}} = -\partial_r, Y_{\frac{1}{2}} = -t\partial_r + ir \partial_\xi,  M_0=i\partial_\xi\equiv -{\cal M},
\label{agerealiz}
\eea
where $g(t), h(t)$ are arbitrary time-dependent functions and $\gamma,\delta$ are arbitrary constants. In arriving at (\ref{agerealiz}) we have kept the form of the generators $M_0$ and of the spatial translation $Y_{-\frac{1}{2}}$ and generalized Galilean-invariance $Y_{\frac{1}{2}}$ unchanged. This general realization is central for the new results presented in what follows.

Next we solve the partial differential constraints imposed on the 3-point functions (namely that they are left invariant by the Age generators). The conclusion is that the most general scalar 3-point function is
\bea
G_{A}(\{t_i, r_i, \xi_i\})=\bigg(\prod_{i=1}^3 t_i^{\gamma_i}\bigg)
\exp(\sum_{i=1}^3 \int^{t_i} d\tau\frac{g(\tau)}\tau)
\times
(t_3-t_1)^{-\tfrac 12 \Delta_{31,2}+\gamma_{31,2}} (t_3-t_2)^{-\tfrac 12 \Delta_{32,1}+\gamma_{32,1}} \nn\\
\times
(t_2-t_1)^{-\tfrac 12 \Delta_{21,3}+\gamma_{21,3}}  \Theta_A\bigg(u_1, u_2, u_3, \frac{t_3(t_2-t_1)}{t_2(t_3-t_1)}\bigg)\label{mostgenage}
\eea
where 
$
\gamma_{31,2}=\gamma_3+\gamma_1 -\gamma_2 \;{\rm etc} $ and
$\Delta_{31,2}=\Delta_3+\Delta_1 -\Delta_2 \;{\rm etc}$.
Here $\Theta_A(u_1, u_2, u_3, \frac{t_3(t_2-t_1)}{t_2(t_3-t_1)})$ is some unconstrained function of:
\bea
u_1=-2i(\xi_2-\xi_1) + \frac{(r_2-r_1)^2}{t_2-t_1}+2\int_{t_1}^{t_2} d\tau \frac{h(\tau)}{\tau}
 + 2(\delta_2-\delta_1)\ln(t_2-t_1)\nn
\\+2(\delta_1  +\delta_2) \ln\frac{t_3-t_2}{t_3-t_1} 
-2\delta_2\ln t_2+2\delta_1\ln t_1,\nn\\
 u_2=
  -2i(\xi_3-\xi_1) + \frac{(r_3-r_1)^2}{t_3-t_1}+2\int_{t_1}^{t_3} d\tau \frac{h(\tau)}{\tau}
 + 2(\delta_3-\delta_1)\ln(t_3-t_1) \nn\\ +2(\delta_1+\delta_3) \ln\frac{t_3-t_2}{t_2-t_1}
-2\delta_3\ln t_3+2\delta_1\ln t_1,\nn\\
u_3= -2i(\xi_3-\xi_2) + \frac{(r_3-r_2)^2}{t_3-t_2}+2\int_{t_2}^{t_3} d\tau \frac{h(\tau)}{\tau}
 + 2(\delta_3-\delta_2)\ln(t_3-t_2) \nn\\+2(\delta_2+\delta_3) \ln\frac{t_3-t_1}{t_2-t_1}
-2\delta_3\ln t_3+2\delta_2\ln t_2.
\eea
For the Schr\"odinger 3-point correlator we find a similar expression, but without the dependence on the additional variable
$\frac{t_3(t_2-t_1)}{t_2(t_3-t_1)}$:
\bea
G_{S}(\{t_i, r_i, \xi_i\})=\bigg(\prod_{i=1}^3 t_i^{\gamma_i}\bigg)
\exp(\sum_{i=1}^3 \int^{t_i} d\tau\frac{g(\tau)}\tau)
\times
(t_3-t_1)^{-\tfrac 12 \Delta_{31,2}+\gamma_{31,2}} (t_3-t_2)^{-\tfrac 12 \Delta_{32,1}+\gamma_{32,1}} \nn\\
\times
(t_2-t_1)^{-\tfrac 12 \Delta_{21,3}+\gamma_{21,3}}  \Theta_S(u_1, u_2, u_3).\label{mostgensch}
\eea
Note that despite the presence of the time-dependent prefactors this correlator is time-translation invariant.
In fact, it is easy to check that a redefinition of the primary fields of the Schr\"odinger algebra, effected by factoring out appropriate time-dependent functions, gives the correlators of the type (\ref{mostgensch}).
However, this redefinition does not change the fact that $X_{-1}G_S(\{t_i, r_i, \xi_i\})=0$. The fundamental difference between Age and Schr\"odinger 3-point functions lies in the dependence of the former on
$\frac{t_3(t_2-t_1)}{t_2(t_3-t_1)}$. 

At this stage we pause to note that the analysis reviewed in the previous part of this section regarding the form of the Age correlators was too restrictive.
 
Since the time-dependent potential is introduced by a simple redefinition of the fields, $\varphi_S(t,\vec x)=\exp(i\int^t d\tau v(\tau)) \varphi_A(t, \vec x)$, the relevant symmetry group is still the full Schr\"odinger and {\it not} the Age group. One of the consequences of this observation is that the holographic realization of aging reviewed in the previous part of this section, is equally restrictive, and thus, the most general holographic Age background is yet to be found. As noted in Ref. \refcite{age3} this is further evidenced by the fact that the 3-point correlators implied by the aging metric found in Ref. \refcite{age2} are dressed Schr\"odinger correlators (i.e. they are ``fake'' Age correlators), whereas the ones in (\ref{mostgenage}) are not.

Finally, following Ref. \refcite{age3} we comment on the holographic realization of the Age algebra in terms of metric isometries of a 1+3-dimensional space. The holographic dual space is parametrized by $x^\mu$ coordinates: $t, r ,\xi$ and the holographic coordinate $z$. The main observation is that once one identifies the Killing vectors $K=K^\mu \frac{\partial}{\partial x^\mu} $ obeying the Age algebra, one can reverse engineer the metric by solving the Killing vector equations for the components of the metric, i.e. $(g_{\rho\nu} \partial_\mu + g_{\rho\mu}\partial_\nu)K^\rho+K^\rho \partial_\rho g_{\mu\nu}=0$. It is natural to assume that $Y_{-\frac{1}{2}}$ and $M_0$ are bulk Killing vectors. If one makes the additional assumption that $Y_{\frac{1}{2}}$, given by (\ref{agerealiz}), is a bulk Killing vector then the problem becomes quite tractable. The bulk forms of the Killing vectors $X_{0}$ and $X_1$ are
$
X_0=-t\partial_t+X_0^\xi \partial_\xi -\tfrac 12 r\partial r 
+X_0^z\partial_z 
$
and
$
X_1=-t^2\partial_t+X_1^\xi\partial_\xi -tr\partial_r +X_1^z\partial_z,
$
where
$X_0^z=\frac{\partial_t (tg_{rr})}{\partial_z g_{rr}}$,
$X_1^z=\frac{\partial_t(t^2 g_{rr})}{\partial_z g_{rr}}$ and
\bea
X_0^\xi =\frac{i}{2t\partial_z g_{rr}}\bigg
(-\partial_z(g_{rr} S)-t\partial_t g_{rr} \partial_z S
+t \partial_t S\partial_z g_{rr} 
+ 2tT\partial_z g_{rr}  + 2tC_1\partial_z g_{rr}\bigg) \nn\\
 X^\xi_1=\frac{i}{2\partial_z g_{rr}}\bigg(-z^2 \partial_z g_{rr}-2 g_{rr}\partial_z S - S\partial_z g_{rr}-t\partial_t g_{rr}\partial_z S 
+ t\partial_t S\partial_z g_{rr}+2tT \partial_z g_{rr} \bigg).
\eea
Here $g_{rr}=g_{rr}(t,z), S=S(t,z), T=T(t)$ and $C_1$ is an arbitrary constant. Solving the Killing vector equations corresponding to $Y_{-\frac{1}{2}}$ and $M_0$ leads to a metric which is $\xi, r$-independent. Furthermore,  solving the $Y_{\frac{1}{2}}$ Killing equations brings the metric to a form which coincides with the initial ansatz of Ref. \refcite{age2}:
$
ds^2 = g_{tt}(t,z) dt^2 + g_{rr}(t,z) dr^2 + g_{zz}(t,z) dz^2 
- 2i g_{rr}(t,z) dt d\xi + 2 g_{tz}(t,z) dt dz.
$
The other components of the metric are determined by the remaining Killing equations:
$g_{zz}=\frac{C_2 (\partial_z g_{rr})^2}{g_{rr}^2} $ and
\bea
g_{tz}=\frac{g_{rr}\partial_z S}{2t}+\frac{C_2 \partial_z g_{rr}\partial_t(t^2 g_{rr}) }{t^2 g_{rr}^2} + C_1 \partial_z g_{rr}\nn\\
g_{tt}= C_3 g_{rr}^2 + \frac{C_2(\partial_t(t^2 g_{rr}))^2}{t^4g_{rr}^2}
+
\frac{2C_1\partial_t(t^2 g_{rr})}{t^2}+\frac{g_{rr}(2tT-S+t\partial_tS)}{t^2},
\eea
\noindent where $C_2, C_3$ are additional integration constants. We stress that this metric is the {\it most general} solution of the reverse-engineering procedure, given the confines of the initial assumption that $Y_{\frac{1}{2}}$ becomes a bulk Killing vector while remaining unchanged.
However, we cannot claim that we have identified the holographic dual of a general theory possessing the full symmetry of the Age algebra. The reason for this is that we are able to identify one more Killing vector of the metric compatible with the following bulk extension of $X_{-1}$
\begin{equation}
-X_{-1}=\partial_t  
 - \frac{\partial_t g_{rr}}{\partial_z g_{rr}}\partial_z
+i\bigg(-\frac{2 C_2}{t^2 g_{rr}}+ \frac{S}{2t^2} - \frac{\partial_t S+2T +4C_1}{2t} + \frac{\partial_t g_{rr}\partial_z S}{2t \partial_z g_{rr}}  
\bigg)\partial_\xi .
\end{equation}
Thus the isometries of the above metric generate the full Schr\"{o}dinger algebra, as in Ref. \refcite{age2}. Naturally the correlators computed from this metric using holography exhibit the kind of "fake" aging discussed earlier, and are constrained by the full Schr\"odinger algebra.
For the time being we can only trace this feature to the assumption made regarding the bulk realization of $Y_{\frac{1}{2}}$. (This assumption was also made in Ref. \refcite{age2}.) Relaxing this condition makes the problem of identifying the holographic metric of Aging much more complicated. This is still an important open problem.

\section{The Jarzynski Identity and the AdS/CFT Duality} %

In this section, following Ref. \refcite{jarzads}, we review a deep analogy between the Jarzynski identity \cite{Jar97,jarz2},
one of the most remarkable results in the recent history of non-equilibrium statistical physics,
and the AdS/CFT duality \cite{adscft1,adscft2,adscft3}, one of the most influential developments in
the recent history of quantum field theory and string theory.
The Jarzynski identity has been tested in many experimental situations
in non-equilibrium systems \cite{exp1,exp2,exp3} 
and it has been also theoretically generalized \cite{Hat01,general2a,general2b,general2c}.
On the other hand, the AdS/CFT duality has been used in fields as diverse as
quantum gravity, quantum chromodynamics, nuclear physics, and condensed
matter physics \cite{Son:2008ye,Balasubramanian:2008dm,adscftapp1,adscftapp2,adscftapp3,adscftapp4}. 

The Jarzynski identity \cite{Jar97,jarz2} gives the {\it exact} relation between 
the thermodynamic free energy differences $\Delta G$ and the irreversible work $W$
\be
\langle \exp{(- \beta W)} \rangle = \exp{(- \beta \Delta G)}\label{freeG}
\ee
where $\beta^{-1} = k_B T$ with $k_B$ denoting the Boltzmann constant and 
$T$ the temperature.
The average $\langle...\rangle$ is over all trajectories that take the system from
an initial to the final equilibrium states.
Note that this exact equality extends the well known inequality between
work and change in free energy, $W \ge \Delta G$,  which follows from the second law
of thermodynamics. The relation  $W \ge \Delta G$ is implied by the
Jarzynski identity and Jensen's inequality $\langle e^A\rangle \ge e^{\langle A\rangle}$.

In the AdS/CFT correspondence, one computes the on-shell
bulk action $S_{bulk}$ and relates it to the appropriate boundary
correlators. The {\it conjecture} \cite{adscft1,adscft2,adscft3} is then that
the generating functional of the vacuum correlators of
the operator $O$ for
a $d$ dimensional conformal field theory (CFT) is given by the
partition function $Z(\phi)$ in (Anti-de-Sitter) $AdS_{d+1}$ space
\be
\langle \exp(\int JO ) \rangle = Z (\phi) \to
\exp[- S_{bulk}( {g}, \phi,...)].\label{part}
\ee
where in the semiclassical limit the
partition function $Z= \exp(-S_{bulk})$.
Here $g$ denotes the metric of the $AdS_{d+1}$ space,
and the boundary values of the bulk field $\phi$ are given by
the sources $J$ of the boundary CFT. Essentially here one reinterprets the RG flow of the
boundary non-gravitational theory in terms of
bulk gravitational equations of motion, and then
rewrites the generating functional of vacuum correlators
of the boundary theory in terms of a semi-classical
wave function of the bulk ``universe'' with specific boundary
conditions. Note that we have
written here a semi-classical expression for the correspondence,
which is what is essentially used in many tests of this remarkable
conjecture \cite{Son:2008ye,Balasubramanian:2008dm,adscftapp1,adscftapp2,adscftapp3,adscftapp4}.

Obviously, there exists a naive formal similarity between the expressions (\ref{freeG}) and (\ref{part}),
given the fact that $\int JO$ formally corresponds to generalized ``work''.
What was argued in Ref. \refcite{jarzads} is that this naive similarity is actually deeper
and points to a profound analogy between the two relations.
Given the fact that (\ref{freeG}) is exact (under certain assumptions) and (\ref{part}) is
still regarded as conjectural, but extremely profound and technically powerful,
this analogy might point a way for a formal ``proof'' of (\ref{part}).
Also, this analogy points to some novel views on the
RG flows of quantum field theories as well as to natural generalizations of
the AdS/CFT dictionary.

The precise definition of $W$ and $G$ is as follows \cite{jarzads}: Given the canonical Liouville equation whose stationary solution is the
Boltzmann distribution $e^{-\beta H (\vec{x},t)}$, one can define \cite{szabo} the average over an ensemble of trajectories starting from the
equilibrium Boltzmann distribution at $t=0$ and evolving according to the Liouville equation. Each trajectory is weighted with the Boltzmann factor of the external work $W_t$ done
on the system\cite{szabo}
\be
W_t = \int_0^t \frac{\partial H}{\partial t'} (\vec{x_t'}, t') dt'.
\ee
By remembering that the exponent of the free energy difference is given by definition as
\be
e^{-\beta \Delta G} \equiv \frac{\int d \vec{x} e^{-\beta H(\vec{x}, t) } }{\int d\vec{y} e^{-\beta H(\vec{y}, 0)} }
\ee
and by using the results of Ref. \refcite{szabo} one is lead to the Jarzynski identity
\be
\exp{(-\beta \Delta G)} \equiv \frac{\int d \vec{x} e^{-\beta H(\vec{x}, t) } }{\int d\vec{y} e^{-\beta H(\vec{y}, 0)} } = \langle \exp{(- \beta W_t)} \rangle .
\ee

In Ref. \refcite{jarzads}, this procedure was applied to the RG trajectories, by using the 
well-known formal dictionary between the Hamiltonian $H$ of a dynamical system in phase space
and the action $S$ of a Euclidean quantum field theory \cite{wilson1,wilson2} 
\be
\beta H(\vec{x}, t) \to S(\varphi, \Lambda)
\ee
where $\Lambda$ denotes the cut-off at which the action of
the quantum field theory is evaluated according to the dynamical RG equation \cite{wilson1,wilson2}.
The RG evolution parameter (``RG time'') is given by the fact that the operation of
rescaling formally corresponds to the ``temporal'' evolution
\be
\Lambda \frac{ \partial}{\partial \Lambda} \to \frac{\partial }{\partial \tau}.
\ee
In Ref. \refcite{jarzads} a Jarzynski-like identity was derived in the context of the renormalization group
by following the proof \cite{szabo} of the Jarzynski identity for the case of the Liouville dynamics.
Note that the renormalization group
dynamics is {\it not} the Liouville dynamics, because of its fundamental
irreversible nature and yet the {\it logic} applied to the context
of the Liouville dynamics can be used in the context of the renormalization
group in order to arrive at the statement of the Jarzynski-like
identity! The important point here is that in the context of
the Wilsonian renormalization group one ultimately gets a stochastic like
equation which is then solved by averaging over the renormalization group
trajectories for the appropriate expressions involving the ``free energy''
and ``work''. This then leads to a new Jarzynski-like identity
involving averages over ensembles of RG (and not dynamical) trajectories!
Here one should emphasize that both the RG ``free energy''
and ``work'' introduced below are defined with respect
to the renormalization group formalism and are fully covariant. 

More precisely, each RG trajectory is weighted with the appropriate ``Boltzmann factor'' of the ``external work'' $W_{\tau}$ done
on the system
\be
W_{\tau}= \int_0^\tau \frac{\partial S}{\partial \tau'} (\varphi_{\tau'}, \tau') d\tau'~.
\ee
Also, the ``free energy'' difference is given by definition as
\be
e^{-\Delta G} \equiv \frac{\int D \varphi e^{-S(\varphi, \tau) }}{\int D \psi e^{-S(\psi, \tau_0)} }
\ee
where $\tau_0$ corresponds to the initial cutoff $\Lambda_0$.
We are thus lead to the RG form of the Jarzynski identity
\be \label{RGJ}
\exp{(- \Delta G)} \equiv \frac{\int D \varphi e^{-S(\varphi, \tau)} }{\int D \psi e^{-S(\psi, \tau_0)} } = \langle \exp{(- W_\tau)} \rangle .
\ee
This equations has not been considered in the literature before, even though
it is of an exact form, presumably because the physical construction that leads to
the relevant linear stochastic equation which implies this exact equality,
is not really motivated without thinking about the original Jarzynski equality.

Next, we equate the work $W_{\tau}$ with the work done by the external source.
This can be understood very simply by invoking the conjugate relation between the
sources and fields with respect to the covariant action. The relation of this
type defines generalized forces (in this cases, sources) and thus 
\be
W_{\tau} \equiv  - \int J \varphi 
\ee
can be understood as a generalized work (where the integral is over space). 
We think that this substitution is natural given the covariant nature
of the RG Jarzynski identity, and the fact that in the case of vacuum
averages, which we have argued replace the average over the RG trajectories, the only ``covariant work'' is done by sourcing the vacuum. We do not see any other natural candidate for such
``covariant work''. 
We also identify the initial and 
the final conformal fixed point and apply the above proposal for the
Jarzynski-like identity in the renormalizaton group context and then we are led to an AdS/CFT-like relation
\be
\langle \exp(\int J \varphi ) \rangle =
\exp (- \Delta G ).
\ee
Note, that we have treated $\varphi$ as the fundamental field.
The same reasoning can be applied to any general operator $O$
in the above Euclidean quantum field theory.
Now, we would like to appeal to the extra dimension $\tau$ to argue
that this formula can be rewritten as the actual AdS/CFT relation
provided:

1) We assume a geometrization of assumed conformal invariance in the $\tau$ direction, so
that the metric in the $\tau$ direction has the isometries of the conformal group associated
with the assumed initial and final conformal fixed points.
This leads us to asymptotically AdS metrics
$ds^2 = d r^2 + e^{A r} ds^2_{CFT}$, where $\tau = A r$ in the flat coordinate system 
($A$ determining the size of the bulk space) and
where $ds^2_{CFT}$ is the natural flat metric of the boundary CFT.

2) We assume a map between the choice of RG scheme to the choice of coordinates
in the $\tau$ extra dimensional space, thus effectively inducing gravitational
interactions in this AdS space. This is reasonable from what we know about
perturbative string theory and its relation to the Wilsonian RG \cite{stringrg1,stringrg2,stringrg3}, as well
as from what we know about holographic RG in the context of AdS/CFT \cite{holorg1,holorg2,holorg3,holorg4,holorg5,holorg55,holorg6,holorg7}.
Nevertheless, this might be harder to justify than the first assumption.

As another general caveat we note that the field theories for which
we expect holographic duals are gauge theories for which we do not have
a nice Wilsonian RG because a cutoff corresponding to a physical length
scale typically breaks gauge invariance, and a cutoff for the gauge theory that could be geometrized is not known at present. 
Also, most field theories do not have a semi-classical
gravity dual and thus AdS/CFT should work only for a limited number
of quantum field theories. Most probably, the theories for which this
duality works can be obtained from the open string sector, in which
case AdS/CFT is really an open/closed string duality of a very specific kind (the gravity dual coming from the closed string sector). A nice discussion of this point is given by
I. Heemskerk and J. Polchinski in the first reference of Refs. \refcite{joesb,joesb1,joesb2,joesb3,joesb4}.

Finally, we recall that gravity is a very special interaction whose energy is given in term of boundary data \cite{marolf1,marolf2,marolf3}, or symbolically
\be
\Delta G = \Delta S_{bulk}
\ee
and thus the RG Jarzynski identity, with above assumptions, becomes the canonical AdS/CFT formula.
Note that the semiclassical limit has to come in here, if the expression for the change of the free energy defined in the context of the RG Jarzynski identity is used so that the relative partition function is expanded in some appropriate WKB limit. That WKB expression for the
relative partition function will necessarily involve an exponent of some effective action,
which could be interpreted as an on-shell ``bulk'' action.
Of course, the reason for the fundamental appearance of gravity is obscure in this
heuristic argument. Presumably the true origin of gravity
in the AdS/CFT duality should be sought in the open/closed string duality.

Next, we can try to apply the generalizations of Jarzynski's
identity \cite{Hat01} in order to generalize AdS/CFT-like dualities.
In that case we do not need to assume conformally invariant fixed points.
On the side of non-equilibrium physics \cite{Hat01},
this would correspond to the situation where one has a probability distribution
of a steady state (ss) with some parameter $\alpha$, $\rho_{ss}(x; \alpha)$, with
the corresponding (negative) ``entropy'' (in the sense of Boltzmann's definition)
\be
\Phi(x; \alpha) = -\log{\rho_{ss}(x; \alpha)}~.
\ee
Given the general properties of probability distributions one
can assert the following mathematical identity \cite{Hat01}
(for a discrete time evolution, labeled by $i=1,2,...N$)
\be
\langle \prod_{i=0}^{N-1}\frac{\rho_{ss}(x_{i+1}; \alpha_{i+1})}{\rho_{ss}(x_{i+1}; \alpha_{i})}\rangle =1
\ee
that implies in the limit $N \to \infty$ \cite{Hat01} the generalized Jarzynski identity
\be
\langle \exp(-\int_0^t d t' \frac{d \alpha}{d t'} \frac{\partial \Phi(x;\alpha)}{\partial \alpha})\rangle =1.
\ee
The usual Jarzynski identity follows when $\Phi = -\beta (G-W)$.
Given our dictionary between time and the logarithm of the cut-off  $\Lambda$
($ t \to \tau$) we can obviously translate this general Jarzynski formula into
a general AdS/CFT-like formula, 
\be
\langle \exp(-\int_0^\tau d \tau' \frac{d \alpha}{d \tau'} \frac{\partial \tilde{\Phi}(x;\alpha)}{\partial \alpha})\rangle =1
\ee
which, curiously, involves the gradient of 
``entropy`` $\frac{\partial \tilde{\Phi}}{\partial \alpha}$.
(In the usual AdS/CFT case $\tilde{\Phi} = -(S_{bulk}+\int J O)$.)
This gradient of ``entropy'' corresponds to some kind of ``entropic force'',
a concept that has recently been invoked in the context of the holographic
treatment of gravity \cite{verlinde,verlinde1}. Thus, it is quite plausible that the
concept of entropic force does play a very precise, albeit hidden, role in the AdS/CFT-like
dualities. Such a generalized AdS/CFT formula should be useful in
illuminating the puzzling duals of cosmological backgrounds or pure (non-conformal)
Yang-Mills theory, or various condensed matter systems.

\section{Turbulence, Emergent Gauge Symmetries and Strings} %

In this section we review the proposal \cite{dmturb} for universal steady state distributions for fully
developed turbulent flows in two and three dimensions (2d and 3d)
building on the previous work which aims to relate modern quantum field theory, string theory
and fluid dynamics \cite{turb,turb1,turb2}.
The basic dynamical equation for turbulent flow is the non-linear Navier-Stokes equation (we use the sum convention throughout the section)
\begin{equation}
\rho (\partial_t v_i + v_j \partial_j v_i) = - \partial_i p + \nu \partial_j^2 v_i,
\end{equation}
with the incompressibility condition $\partial_i v_i = 0$, where $v_i$ is a component of the velocity field of the flow, $p$ is pressure, and $\rho$ is the fluid density \cite{review}.
In this section we are interested in fully developed
turbulence, or turbulence in the limit of infinite Reynolds number $R$. As $R = L v/{\nu}$
goes to infinity, with $v = \sqrt{v_i v_i}$, whereas $L$ is a characteristic scale and $\nu$ is the kinematic viscosity, we effectively have the limit of vanishing
viscosity $\nu \to 0$. In this regime, all the various possible symmetries are restored in a statistical sense,
calling for a probabilistic description of what is in essence a deterministic system (strongly dependent on the boundary conditions). Therefore,
in computing correlators of the velocity field, we should use the statistical and
quantum field theoretic descriptions of turbulence \cite{qft,qft1,qft2}.

In particular we are interested in the form of a steady state turbulent distribution $P(\vec{v})$
which should imply the various famous scaling laws (the Komogorov scaling in 3d \cite{ko} and 
both Kolmogorov and Kraichnan scaling laws in 2d \cite{Kra80}). The two-point correlators determined
by these scalings should be computable from $P(\vec{v})$ as follows
\begin{equation}
\langle \delta v_i(l) \delta v_j (0) \rangle \equiv 
\frac{\int D \vec{v} \delta v_i(l) \delta v_j(0) P(\vec{v})}{\int D \vec{v} P(\vec{v}) }.
\end{equation}
where $\delta v_i(l)$ is the $i$-th component of the difference vector
$\vec{v}(\vec{x}+\vec{l}) - \vec{v}(\vec{x})$, with $l=|\vec{l}|$ being the distance between the
fluid elements located at positions $\vec{x}$ and $\vec{x} + \vec{l}$.
The fundamental question is: what is $P(\vec{v})$?
In Ref. \refcite{dmturb} it was argued that the answer to this question is:
\begin{equation}
P(\vec{v}) = \exp[-S_{K} (\vec{v})],
\end{equation}
where we assume that the expression for the $S_K(\vec{v})$ is local and universal (see below). Both of these 
assumptions might be challenged:
we can a priori expect non-local factors in $S_K(\vec{v})$. 
Also, the relative locations of the Kolmogorov and Kraichnan distributions in 2d turbulence energy spectra
depend on the (location of the) forcing, which challenges universality.
We also note that the notion of universality might be challenged by an a priori dependence
on the boundary conditions. We will simply assume that universal distributions exist
in the rest of this section, in spite of these caveats.
Finally, we note that if universality applies to turbulence, it is expected at short 
distance \cite{qft,qft1,qft2}, as opposed to the usual long distance universality associated with
quantum field theory and critical phenomena from equilibrium physics.

The central message is that
$S_K$ is what we call the Kolmogorov distribution in 3d,
determined by the effective action for a 3d gauge theory
based on volume preserving diffeomorphisms. Alternatively,
$S_{K}$ is what we call the Kraichnan distribution in 2d, 
determined by the effective action for a 2d gauge theory based on area preserving diffeomorphisms.
In both cases the diffeomorphisms act in velocity space as emergent symmetries.
In both cases we crucially use the Galilean symmetry in the space of velocities, so that
the effective respective actions involve only derivatives in $v_i$.
We pause briefly to remark that in order to fully characterize systems far from equilibrium
knowledge of the probability currents is needed in addition to the steady state probability 
distributions \cite{zia,zia1}. However, a discussion of these currents is outside of the scope
of the present review.

Even though the puzzle of fully developed turbulence is essentially a strongly coupled
problem, we motivate our discussion (and ultimately, our proposals) by some rather elementary observations at
weak coupling.
We start with 2d and then move on to 3d.
In 2d the Lagrangian  description of the fluid is generated by 
the following Lagrangian \cite{lenny} (in the notation of Ref. \refcite{lenny})
\begin{equation}
L_2= \int d^2y \rho_0 [ \frac{m}{2} \dot{x}^2 - V( \rho_0 |\frac{ \partial y}{\partial x}|)].
\end{equation}
where $\rho_0$ is the constant density in the co-moving coordinates
and the real space density is $\rho_0 |\frac{ \partial y}{\partial x}|$,
where $|\frac{ \partial y}{\partial x}|$ denotes the Jacobian connecting the 
co-moving coordinates $y$ and the continuum fields $x_i(y, t)$.
The Lagrangian $L_2$ is invariant under area preserving diffeomorphisms in 2d \cite{lenny}:
\begin{equation}
y_i' = y_i +f_i(y) \quad;\quad \delta x_a = \frac{\partial x_a}{\partial y_i} f_i(y),
\end{equation}
where for area preserving diffeomorphisms ($\epsilon_{ij}$ being the Levi-Civita symbol in 2d)
\begin{equation}
f_i = \epsilon_{ij} \frac{\partial \Lambda(y)}{\partial y_j},
\end{equation}
and where $\Lambda(y)$ is an arbitrary function, which generates these ``gauge''
transformations. This equation in turn
leads to the 2d Poisson bracket action:
\begin{equation}
\delta x_a = \frac{\partial x_a}{\partial y_i}\epsilon_{ij} \frac{\partial \Lambda(y)}{\partial y_j}
\equiv \{\Lambda, x_a\}~.
\end{equation}
Suppose we consider small motions of the 2d fluid \cite{lenny}:
\begin{equation}
x_i = y_i + e \epsilon_{ij} A_j,
\end{equation}
where $e$ is the small coupling (and where we have absorbed the factor of $\rho_0$ compared to
Ref. \refcite{lenny}). To linear order the area preserving diffeomorphisms become the usual gauge transformations
\begin{equation}
\delta A_i = \partial_i \Lambda,
\end{equation}
and the quadratic Lagrangian that is invariant under this transformation is
just the usual Maxwell Lagrangian
\begin{equation}
L_2 \sim \frac{1}{2 e^2} \int d^2y [ \dot{A_i}^2 - (\nabla \times \vec{A})^2].
\end{equation}
The inclusion of non-linear terms can be done by extending the linear gauge transformations
to their non-Abelian completion. This leads to a non-Abelian gauge theory,
with the following map between the full area preserving diffeomorphism group and the non-Abelian transformations
generated by a commutator of two matrices
\begin{equation}
\{\Lambda, x\} \to [\lambda, X].
\end{equation}
Here we have used the standard \cite{apd,apd1,apd2,apd3,apd4} mapping between
the Poisson brackets and commutators of infinite square matrices $\lambda$ and $X$.
The corresponding gauge theory action reads as
\begin{equation} \label{2dYM}
S_A=  \frac{1}{2 e^2} \int d^2 y [  \dot{A_i}^2 - (\partial_i A_j - \partial_j A_i + [A_i, A_j])^2].
\end{equation}
We will return to this gauge theory action in what follows.
In 3d we can extend the presentation of Ref. \refcite{lenny} by considering
\begin{equation}
L_3= \int d^3y \rho_0 [ \frac{m}{2} \dot{x}^2 - V( \rho_0 |\frac{ \partial y}{\partial x}|)].
\end{equation}
The Lagrangian $L_3$ is invariant under volume preserving diffeomorphisms
\begin{equation}
\delta x_a = \frac{\partial x_a}{\partial y_i}\epsilon_{ijk} \frac{\partial \Lambda_1(y)}{\partial y_j} 
\frac{\partial \Lambda_2(y)}{\partial y_k}.
\end{equation}
where $\Lambda_1$ and $\Lambda_2$ are the generators of the volume preserving
gauge transformations.
This is equivalent to the following Nambu bracket \cite{nambu,nambu1,nambu2}:
\begin{equation}
\delta x_a \equiv \{\Lambda_1, \Lambda_2, x_a\},
\end{equation}
where, by definition 
\begin{equation}
\{A, B, C\}\equiv \epsilon_{abc} \partial_a A \partial_b B \partial_c C.
\end{equation}
Here $A,B,C$ are three functions of three spatial coordinates $x,y,z$.
This classical bracket seems to be naturally generalized to a triple algebraic structure \cite{nambu,nambu1,nambu2,triples,vector,vector1,vector2,vector3,vector4}
\begin{equation}
[A_i, A_j, A_k] \equiv \epsilon_{abc} A_a A_b A_c.
\end{equation}
Suppose we consider small motions of the 3d fluid in analogy with the 2d case:
\begin{equation}
x_i = y_i + f \epsilon_{ijk} B_{jk},
\end{equation}
where $B_{jk}= - B_{kj}$ and $f$ denotes a small coupling.
The ``quantization'' of the volume preserving diffeomorphisms is a 
more involved problem \cite{nambu,nambu1,nambu2,triples}. Nevertheless, even in this case
we expect
a mapping between the full volume preserving diffeomorphism group and the 3-bracket
\begin{equation}
\{\Lambda_1, \Lambda_2, x\} \to [\lambda_1, \lambda_2, X],
\end{equation}
where $\lambda_1$ and $\lambda_2$ are the appropriate matrix realizations of
$\Lambda_1$ and $\Lambda_2$ \cite{triples}.
The 3d action invariant under the linear part of this transformation is
\begin{equation}
S_B = \frac{1}{2 f^2}  \int d^3y [  \dot{B}_{ij}^2 - (\partial_i B_{jk} +\partial_j B_{ki} +\partial_k B_{ij})^2].
\end{equation}
Note that the explicit linear volume preserving transformations are
\begin{equation}
\delta B_{ij} = \partial_i \Lambda_1 \partial_j \Lambda_2 - \partial_i \Lambda_2
\partial_j \Lambda_1,
\end{equation}
where $B_{ij}$  is dual to a 3-vector in three dimensions,
$
a_i = \frac{1}{2} \epsilon_{ijk} B_{jk}
$.

The above linear analyses seem removed from such
a strongly coupled problem as fully developed turbulence \cite{zakharov}. 
Still the linear analysis is useful, because in the stationary case we can formally replace
$x_i \to v_i$ and talk about velocity Lagrangians.
That the corresponding Lagrangians should be given in terms of the derivatives of velocity is
fixed by the Galilean symmetry, $v_i \to v_i + u_i$, for a fixed velocity with component $u_i$.
{\it Moreover, if $v \sim l^\alpha$, from $\delta l = \{F, l\}$ it follows that
we have an emergent symmetry in the space of velocities $\delta v = \{F, v\}$.}
Given this emergent symmetry in 2d we have the following natural theory
that is consistent with area preserving diffeomorphisms involving $v_i$:
\begin{equation}
S^{(2)}_K=  \frac{1}{2 g^2} \int d^2 y [  (\partial_i v_j - \partial_j v_i + t_0 \{v_i, v_j\})^2],
\end{equation}
where $g^2$ and $t_0$ denote the appropriate dimensionful coupling constants and $i,j=1,2$ (the factor $ \frac{1}{2 g^2}$ is canonical).
For $t_0 =1$ this theory is invariant under 
\begin{equation}
\delta v_i = \{ F, v_i \},
\end{equation}
where $F$ generates area preserving gauge transformations in $v_i$ space.
(This is an emergent gauge symmetry in the space of velocities, which should be only the feature
of the scaling regime that characterizes fully developed turbulence.)
We want to argue that this is the stationary distribution we have been
looking for in 2d.
First, we attempt to justify this guess on more general grounds:
Obviously we have the Galilean invariance in the inertial range
$v_i \to v_i + u_i$ (where $u_i$ is constant)
because $S^{(2)}_K$ is a functional of the derivatives of $v_i$.
Second, it is reasonable to expect that $S^{(2)}_K$ is governed by the conserved quantities.
(Recall the case of equilibrium statistical mechanics, i.e. the Boltzmann-Gibbs distribution, where $S$ is simply the energy, an
additive conserved quantity).
In our situation we have vorticity ($\vec{\omega} = \nabla \times \vec{v}$) squared and velocity squared, as the natural
conserved quantities \cite{Kra80, review, ko}.
Therefore, if we start with the vorticity squared ($\omega^2$) term we see that this is
really the quadratic part of our guess for $S^{(2)}_K$. This term is invariant under
the linear gauge transformations in the space of velocities.
However, by going to real space, we may invoke the full non-linear
group of area preserving coordinate transformations generated by
the Jacobian (the Poisson bracket) in 2d space. Then by concentrating on the steady state
regime we may claim the same symmetry in the $v$ space which would
lead us to the above proposal.

This action should be compared to the 2d Yang-Mills theory action $S_A$ given in equation (\ref{2dYM}).
We can use the covariant derivative
$
\partial_i \to D_i \equiv \partial_i + A_i,
$
to rewrite the usual 2d Yang-Mills theory Lagrangian
as
$[D_i, D_j]^2 \equiv (\partial_i A_j - \partial_j A_i + [A_i, A_j])^2$.
This procedure can be immediately generalized to 3d so that we have
a theory consistent with volume preserving diffeomorphisms in the space of $v_i$.
By using the covariant derivative
$\partial_i \to D^v_i \equiv \partial_i + v_i$,
we can immediately rewrite the above guess for the 2d action 
\begin{equation}
S^{(2)}_K=  \frac{1}{2 g^2} \int d^3 y [  \{D^v_i, D^v_j\}^2],
\end{equation}
and then extrapolate our 2d proposal to the natural proposal for the Kolmogorov distribution in 3d
\begin{equation}
S^{(3)}_K=  \frac{1}{2 \tilde{g}^2} \int d^3 y [  \{D^v_i, D^v_j, D^v_k\}^2].
\end{equation}
Here $\tilde{g}^2$ denotes the appropriate dimensionful coupling constant.
The crucial non-linear part
which replaces $\int d^2 y (\{v_i, v_j\}^2)$ is
\begin{equation}
S^{(3)}_K \sim \frac{1}{2 \tilde{g}^2} \int d^3 y [  \{v_i, v_j, v_k\}^2].
\end{equation}
This action is fixed now by the volume preserving diffeomorphisms in the 3d velocity space
\begin{equation}
\delta v_i = \{ F_1, F_2, v_i \},
\end{equation}
where $F_1$ and $F_2$ generate the volume preserving gauge transformations.
The argument which leads to this proposal is just the repetition of the argument we have
presented for the 2d Kraichnan distribution. 

These are thus our explicit proposals for the turbulent distributions in
2d and 3d: they are encoded in the expressions for $S^{(2)}_K$ and $S^{(3)}_K$.
Do these educated guesses give the correct results?
We concentrate on the case of 2d turbulence.
Let us remember that for the Kraichnan scaling in 2d we
want to derive
$
\langle \delta v_i(l) \delta v_j(0)\rangle \sim l^2 \delta_{ij},
$
(for $t_0=1$, so that the natural dimension factor $l/t_0 \to l$)
and for the Kolmogorov scaling law
$
\langle \delta v_i(l) \delta v_j(0)\rangle \sim l^{\frac{2}{3}} \delta_{ij}.
$
(Again using the natural dimensions).
In order to discuss the validity of our proposal, we turn to the
natural loop variables introduced by Migdal \cite{migdal,migdal1,migdal2}.
The natural loop (closed string) variable, the Migdal loop \cite{migdal,migdal1,migdal2}
\begin{equation}
W_M (C)\equiv \langle \exp(-\frac{1}{\nu} \int_C v_i dx_i) \rangle,
\end{equation}
(where $C$ is a contour and the viscosity $\nu$ plays the role
of an effective $\hbar$ \cite{turb,turb1,turb2,migdal,migdal1,migdal2}) 
allows us to rewrite the Naiver-Stokes equations \cite{migdal,migdal1,migdal2}
as an effective Schrodinger equation 
\begin{equation}
i \nu \partial_t W_M(C) = H_C W_M(C),
\end{equation}
with the appropriate loop equation Hamiltonian $H_C$ \cite{migdal,migdal1,migdal2}.
(The loop equation for turbulence is a direct analog of the well-known loop equation
for the non-Abelian gauge theory, also proposed by Migdal and collaborators.)
In 3d Migdal observed a self-consistent scaling solution of this equation
in the $\nu \to 0$ limit (a WKB limit in this problem)
$
W_M(C) \sim \exp( - [\frac{A}{A_0}]^{\frac{2}{3}}),
$
which precisely corresponds to the Kolmogorov scaling.
In 2d, the Kraichnan scaling leads to
the area law \cite{turb,turb1,turb2} for the Migdal loop
$
W_M(C) \sim \exp( -A/A_0).
$
In the
above gauge theory of 2d velocities this area law is very natural,
because of the fact that in the corresponding 2d Yang-Mills theory, the Wilson loop (the natural
analog of the Migdal loop)
\begin{equation}
W(C)\equiv \langle \exp(-\int_C A_i dx_i) \rangle,
\end{equation}
obeys the same area law \cite{wilson}
$
W(C) \sim \exp(-A/A_0).
$
Given the precise structural mapping between the 2d Kraichnan theory 
(defined by $S^{(2)}_K$) and
the 2d gauge theory, this area law scaling for the loop variables is obeyed, and the Kraichnan scaling
thus follows from our proposed steady state distribution.
This has been checked in unpublished numerical experiments.\cite{unpublished}

\section{Conclusion: Quantum Gravity vs. Non-equilibrium Physics}

In this concluding section of the review we want to argue for a more general relation of
quantum gravity and non-equilibrium physics, as presented in Refs. \refcite{time,dmreview}.
First we summarize some facts known in the theory of dynamical systems \cite{ruelle,dorfman}.
Start with a non-linear Hamiltonian system with slight dissipation (see below).
Then:


%

1) Non-linearities generate positive dynamical Lyapunov exponents which ultimately lead
to chaotic dynamics (the negative Lyapunov exponents are irrelevant on the long time scales).
The chaotic dynamics manifests itself in the emergence of the attractor.
There exist natural measures on this attractor, the famous Bowen-Ruelle-Sinai measure (BRS). \cite{ruelle,dorfman}
The relevant equations here are as follows. \cite{ruelle,dorfman}
One starts from {\it dissipative} dynamics
\begin{equation}
{d p_a \over dt} = - \kappa \frac{\partial F}{\partial p_a}, \quad {d q_a \over dt} = p_a.
\end{equation}
The dissipative term, described by the dissipative functional $F$,
for example
\be
F \sim \sum_{ij} a_{ij} p^i p^j
\ee
is generated by looking at an infinite system
because essentially an infinite reservoir can be understood as a thermostat.
Then one integrates out the reservoir degrees of freedom to produce a
finite non-Hamiltonian (dissipative) system.
Because the dynamics is non-Hamiltonian, the volume of phase space {\it is not} preserved!
The entropy production is, at the end, crucially related to 
the volume construction of the phase space volume.
Note that it is important here that there is hyperbolic dynamics (i.e. chaos) in the effective dynamical
equations
\be
\frac{dx}{dt} = g(x)
\ee
so that the time evolution $x(t) = f^t x(0)$
does have positive Lyapunov exponents $x(t) \sim e^{l t}$, with $l >0$.
This leads to chaotic time evolution and the stretching of some directions of a
unit phase space volume.
Finally, there are measures that are invariant under time evolution.
The BRS measure $\rho$ may be singular (it is defined on the attractor which is usually a
fractal space) but the time evolution averages are indeed averages over this measure, so that
\be
\lim_{T \to \infty} \frac{1}{T +t} \int_{-T}^{t} d\tau O(f^{T +\tau} x) = \int \rho (dy) O(y)
\ee
This is the crucial relation which calls for a holographic interpretation in the gravitational context \cite{time}.
Notice that the dynamics is that of a $d+1$ dimensional system (in this case 
the extra dimension $t$ is 
indeed a time parameter, but it could be some other evolution parameter such as 
the radial slicing of $AdS$ space), that is related to an 
ensemble description of a $d$ dimensional system, with a specific (BRS) measure.

2) A crucial aspect of this theory is entropy production and
volume contraction (note that dissipation is crucial for the appearance of
the attractor and the new measures).
The introduction of the new measures is tantamount to the breaking of time-reversal
symmetry - which is ultimately the consequence of dissipation.
The central equation for the entropy $S$ in terms of the invariant measure $\rho$ is \cite{ruelle,dorfman}
\be
S(\rho) = \int \rho(dx) ( - \nabla_x g)
\ee
where the divergence is computed with respect to the phase space volume element.
For the BRS measure one can show that $S(\rho) \ge 0$ \cite{ruelle,dorfman}.
Once again, in view of the relation between quantum gravity and non-equilibrium physics argued in
this review, it is tempting to interpret this entropy production precisely as the
gravitational entropy in the gravitational context. 

3) Given the natural dynamically generated measure on the attractor, 
in the near-to-equilibrium limit of the response functions for this general
dynamics one gets an averaged, coarse grained description as required
by the ergodic theorem: i.e. an averaged ensemble description with respect to 
the above measure.
The linear response formulae read as follows \cite{ruelle,dorfman}.
Modify the evolution equation by adding small perturbations
\be
\frac{dx}{dt} = g(x) + X_t(x)
\ee
to compute, via a perturbation in the BRS measure, $\rho \to \rho + \delta_t \rho$
\be
\int \delta_t \rho (dy) O(y) = \int_{-\infty}^{t} d\tau \int \rho (dy)  X_t(y) \nabla_y[O(f^{t -\tau} y)] .
\ee
Note that this should be compared to the usual holographic dictionary for the computations
of various correlations functions.

The obvious application of this set-up is in the context of the AdS/CFT correspondence
by reinterpreting the evolution parameter as the AdS radial ``time''
\be
ds^2 = dr^2 + g_{ij} dx^i dx^j
\ee
and by writing the AdS/CFT dictionary in terms of the above non-equilibrium result,
with $\rho$ being the Euclidean path integral (i.e. the Gibbs measure).
Note that in this case one would consider the boundary of AdS as the attractor for the
bulk quantum gravitational evolution viewed as a dynamical system out of equilibrium.


Similarly, this picture could provide a different view on the issue of cosmological holography.
For the case of time dependent backgrounds the evolution parameter is time-like
as in the context of dynamical systems out of equilibrium
\be
ds^2 = - dt^2 + h_{ij} dx^i dx^j.
\ee
For a de Sitter ($dS$)-like asymptotics the attractor (the holographic screen) does not have to be at all regular.
For example, the $I^{+}$ of eternal inflation looks like a 
fractal. On that fractal-like attractor, we can envision placing a holographic 
dual, given the more general BRS measure\cite{ruelle,dorfman} (and not the Gibbs measure).
One obvious use of this picture is the context of dS holography where it has been conjectured that
a non-standard CFT \cite{vijay}
is dual to an asymptotic dS space, however with a non-standard (non-Gibbsian), BRS measure on the future fractal-like boundary.



Finally, we note that 
given the fact that the Schr\"odinger equation can be viewed as a geodesic equation \cite{time, dmreview}, and given the fact the
the master equation for non-equilibrium physics, such as the Liouville equation, is of the same nature as the Euclidean Schr\"odinger equation, it
is natural to 1) derive the Schr\"odinger and the master equations from some underlying geometric
Einstein equations \cite{time, dmreview} and 2) identify the non-equilibrium equations of this kind with the non-perturbative
formulation of quantum gravity. This has been discussed in Refs. \refcite{time,dmreview,mt,mt1,mt2,jm,jm1,jm2,chia1,chia2}.
According to this proposal the non-perturbative string theory viewed as non-perturbative theory of
quantum gravity and matter is essentially a new generalized quantum theory, which is related to general
non-equilibrium statistical physics, the same way the canonical quantum theory is related to the equilibrium
statistical physics.
(Note that 'tHooft has also emphasized the relation between dissipative non-equilibrium dynamics, holography and quantum gravity
in Ref. \refcite{thooft}.)

Of course, this more general proposal goes beyond what has been reviewed in this article.
Nevertheless, we hope that the material presented in this review
clearly points to a strong relation between quantum gravity, string theory and
non-equilibrium statistical physics that should prove fruitful in finding out new results on both sides of this exciting connection.

\section{Acknowledgements} %
\vskip .5cm
We would like to thank Rob Leigh, Juan Jottar, Leo Pando Zayas, Diana Vaman, Chaolun Wu, Anne Staples,  Vishnu Jejjala, Mike Kavic, Jack Ng and Chia Tze, for many enjoyable and fruitful collaborations. We thank Laurent Freidel for recent discussions on non-equilibrium physics and black hole thermodynamics. D.M. thanks the Aspen Center for Physics, the National Institutes for Health, the Galileo Galilei Institute for Theoretical Physics in Florence, Michigan Center for Theoretical Physics, University of Virginia, International Center for Theoretical Phyics, Trieste, and Perimeter Institute for Hospitality. 
D.M. is supported in part by the US Department of Energy
under contract DE-FG05-92ER40677. 
M.P. is supported by the US National Science Foundation through DMR-0904999 and DMR-1205309.

\vskip .5cm

\appendix

\section{A New Kind of Complementarity}

Finally we make a comment about the not so well known complementarity between
thermodynamics and statistical physics first emphasized by Bohr and worked out 
in a bit more detail by Rosenfeld \cite{br}.

Let us examine the energy-time principle
\be
\delta E \delta t \sim \hbar
\ee
and take the well known relation between the Euclidean time ($\tau = it$) and
the temperature [$\tau = \hbar \beta$, where $\beta = 1/(kT)$].
This relation follows from the formal relation between the Euclidean path integral and
the Boltzmann-Gibbs equilibrium partition function
(and it is the basis for the simplest derivation of the Unruh-Hawking formula
for the uniformly accelerated observer.)
By plugging the relation $\tau = \beta \hbar \equiv \hbar/(kT)$ into the Euclidean energy-time uncertainty
$\delta E \delta \tau \sim \hbar$ we get (if we properly Wick-rotate) the Bohr-Rosenfeld relation
\be
\delta E \delta T \sim k T^2.
\ee
Now, this is curious because (as observed first by Niels Bohr and then by Leon Rosenfeld)
if one were to compute the fluctuations of energy and temperature a la Boltzmann-Einstein,
one would obtain exactly the same formula!
We are not aware that Bohr and Rosenfeld made any connection to the energy-time relation.
They simply observed that from the fluctuation of the internal energy $E$ and temperature $T$ \cite{br}
\be
(\delta E)^2 \sim kT^2 C_V, \quad (\delta T)^2 \sim k T^2/C_V.
\ee
The statistics would enter in the expression for the specific heat $C_V$ at constant volume $V$, 
but upon multiplication the $C_V$ factor cancels. So the final
formula is independent of statistics. The complementarity between the thermodynamic and the statistical descriptions
of macroscopic systems as emphasized by Bohr and by Rosenfeld can be understood as follows. In order to assign a definite temperature to the system it is necessary
to allow the system to exchange energy with a heat bath, and thus, it is impossible
to assign to its energy any definite value. Similarly, in order to keep its energy constant, one must isolate the system, and thus, one cannot assign a definite
temperature to it. \cite{br}

First, we observe that this relation between the fluctuations seems to be the same
as the Euclidean energy time uncertainty plus the relation between the Euclidean
(imaginary) time and temperature.
Second, this relation between Euclidean time and temperature gives us
the relation between temperature and acceleration (the Unruh-Hawking formula)
which then implies
\be
\delta E \delta a \sim \hbar a^2 
\ee
(where we use the Unruh-Hawking relation $T \sim \hbar a)$. 
In the context of string theory 
\be
\alpha' \delta E \sim  \delta x
\ee
and thus one gets, using natural units, that
\be
\delta x \delta a \sim \alpha' a^2
\ee
which points to a complementarity between the position and surface gravity (acceleration).

Given the current discussion on the black hole complementarity \cite{amps,amps1,amps2}, this not-so-well-known complementarity
between thermodynamics and statistical physics, can be immediately extended to the gravitational case,
where the main observation would be that the black hole thermodynamics is complementary to the
statistical description of black holes given by quantum gravity (string theory).

\end{document}